\documentclass[twocolumn,times]{aastex631}

\usepackage{multirow}
\usepackage{amsmath}
\usepackage{enumitem}
\usepackage{gensymb}

\defcitealias{Zhu2023}{Paper I}
\defcitealias{Lu2023}{Paper II}

\hypersetup{linkcolor=blue,citecolor=blue,filecolor=cyan}

\def\equationautorefname~#1\null{Equation~(#1)\null}

\begin{document}
  
\title{MaNGA DynPop. VII. A Unified Bulge-Disk-Halo Model for Explaining Diversity in Circular Velocity Curves of 6000 Spiral and Early-Type Galaxies}

\author[0000-0002-2583-2669]{Kai Zhu}
\affiliation{Department of Astronomy, Tsinghua University, Beijing 100084, People’s Republic of China}
\correspondingauthor{Kai Zhu} 
\email{kaizhu@mail.tsinghua.edu.cn,
zhukai\_astro@outlook.com}
  
\author[0000-0002-1283-8420]{Michele Cappellari}
\affiliation{Sub-department of Astrophysics, Department of Physics, University of Oxford, Denys Wilkinson Building, Keble Road, Oxford, OX1 3RH, UK}

\author[0000-0001-8317-2788]{Shude Mao}
\affiliation{Department of Astronomy, Westlake University, Hangzhou 310030, Zhejiang Province, People’s Republic of China}
\affiliation{Department of Astronomy, Tsinghua University, Beijing 100084, People’s Republic of China}

\author[0000-0002-6726-9499]{Shengdong Lu}
\affiliation{Institute for Computational Cosmology, Department of Physics, University of Durham, South Road, Durham, DH1 3LE, UK}

\author[0000-0003-3899-0612]{Ran Li}
\affiliation{School of Physics and Astronomy, Beijing Normal University,  Beijing 100875, People’s Republic of China}
\affiliation{School of Astronomy and Space Science, University of Chinese Academy of Sciences, Beijing 100049, People’s Republic of China}

\author[0000-0002-8614-6275]{Yong Shi}
\affiliation{School of Astronomy and Space Science, Nanjing University, Nanjing 210093, People’s Republic of China}
\affiliation{Key Laboratory of Modern Astronomy and Astrophysics (Nanjing University), Ministry of Education, Nanjing 210093, People’s Republic of China}

\author[0000-0001-5742-2982]{David A. Simon}
\affiliation{Sub-department of Astrophysics, Department of Physics, University of Oxford, Denys Wilkinson Building, Keble Road, Oxford, OX1 3RH, UK}

\author[0009-0003-3720-6870]{Youquan Fu}
\affiliation{Department of Astronomy, Tsinghua University, Beijing 100084, People’s Republic of China}

\author[0009-0002-8312-1002]{Xiaohan Wang}
\affiliation{Department of Astronomy, Tsinghua University, Beijing 100084, People’s Republic of China}

\received{March 10, 2025}
\revised{July 30, 2025}
\accepted{August 11, 2025}
 
\begin{abstract}

We derive circular velocity curves (CVCs) from stellar dynamical models for $\sim6000$ nearby galaxies in the final data release of the SDSS-IV MaNGA survey with integral-field spectroscopy, exploring connections between the inner gravitational potential (traced by CVC amplitude/shape) and galaxy properties. The maximum circular velocity ($V_{\rm circ}^{\rm max}$) and circular velocity at the half-light radius ($V_{\rm circ}(R_{\rm e}^{\rm maj})$) both scale linearly with the stellar second velocity moment $\sigma_{\rm e}^2\equiv\langle V^2+\sigma^2\rangle$ within the half-light isophote, following $V_{\rm circ}^{\rm max} \approx 1.72\sigma_{\rm e}$ (7\% error) and $V_{\rm circ}(R_{\rm e}^{\rm maj}) \approx 1.62\sigma_{\rm e}$ (7\% error). CVC shapes (rising, flat, declining) correlate strongly with structural and stellar population properties: declining curves dominate in massive, early-type, bulge-dominated galaxies with old, metal-rich stars and early quenching, while rising CVCs prevail in disk-dominated systems with younger stellar populations and ongoing star formation. Using a unified bulge-disk-halo model, we predict CVC shapes with minimal bias, identifying three governing parameters: bulge-to-total mass ratio ($B/T$), dark matter fraction within $R_{\rm e}$, and bulge Sersic index. The distribution of CVC shapes across the mass-size plane reflects evolutionary pathways driven by (i) \emph{in situ} star formation (spurring bulge growth) and (ii) dry mergers. This establishes CVC morphology as a diagnostic for galaxy evolution, linking dynamical signatures to structural and stellar population histories.
\end{abstract}


\section{Introduction} \label{sec:intro}
The rotation curve (RC) of a galaxy represents the rotational velocity of stars or gas as a function of radial distance from the galaxy's center. The first indirect evidence of dark matter (DM) came from studies of the velocity dispersion of galaxies within the Coma Cluster \citep{Zwicky1933,Zwicky1937,Zwicky2009}. Subsequent investigations into DM using galaxy RCs began with nearby galaxies. \citet{Babcock1939} proposed that the nearly flat RC observed at the outskirts of M31 could be explained by an increasing mass-to-light ratio in the outer regions, while \citet{Oort1940} identified substantial amounts of invisible matter in the outer parts of NGC 3115 based on its RC. With the improvements of observational instruments, the RC has long been a powerful tool for studying the mass distributions of spiral galaxies, either using optical emission lines \citep{Rubin1970,Rubin1980,Rubin1982,Rubin1985} or radio 21 cm line \citep{Bosma1979,Bosma1981}.

As indicators of the gravitational potential (encompassing both baryonic and DM mass distributions) within which galaxies reside, the amplitude and the shape of RCs are expected to correlate with various galaxy properties such as luminosity, morphology, and structure. For instance, the Tully-Fisher relation \citep{Tully1977} tightly links the maximum rotation velocities (the amplitudes) and the total luminosities (or stellar masses) of spiral galaxies. Beyond this well-known correlation between RC amplitude and mass, the shapes of RCs have also been found to be dependent on galaxy morphologies \citep{Corradi1990,Erroz-Ferrer2016}, luminosities \citep{Persic1996,Sofue2001}, or a combination of both \citep{Casertano1991,Noordermeer2007,Swaters2009,Yoon2021,Jeong2025}. These evident correlations imply that the gravitational potential plays a crucial role in the evolution of galaxy structure and morphology.

In addition, the gravitational potential also has a significant impact on star-formation quenching, leaving imprints on the present-day stellar population properties. Various mechanisms that halt star formation of galaxies can be categorized into external (outside the galaxies) and internal (within the galaxies). The external mechanisms include ram pressure stripping \citep{Gunn1972}, strangulation \citep{Larson1980}, tidal stripping \citep{Wetzel2013}, dynamical heating from galaxy harassment \citep{Moore1996}, while internal mechanisms include but are not limited to halo quenching \citep{Dekel2006}, active galactic nucleus (AGN) feedback \citep{Harrison2017}, stellar feedback \citep{Colling2018}, mass quenching \citep{Peng2010}, bar quenching \citep{Khoperskov2018}, morphological quenching \citep{Martig2009}, and angular momentum quenching \citep{Lu2022}. Among these quenching mechanisms, the morphological quenching \citep{Martig2009}, also known as gravitational quenching \citep{Genzel2014} or dynamical suppression \citep{Davis2014,Gensior2020,Gensior2021}, occurs when the growth of a stellar spheroid or bulge stabilizes the gas disk instability \citep{Toomre1964}, thereby quenching star formation. This gravitational potential-related mechanism does not require gas removal (e.g. the tidal/ram pressure stripping) or massive halo (e.g. AGN feedback, shock heating induced halo quenching), and thus can explain the appearance of gas-rich but quenched galaxies in halos less massive than $10^{12}{\rm M_{\odot}}$. Linking the gravitational potential and stellar population properties to distinguish gravitational quenching from other quenching mechanisms is crucial in understanding the complex star-formation histories that shape the evolution of galaxies.

The RCs were initially derived from long-slit spectroscopy and only provided one-dimensional information on stellar kinematics or ionized gas kinematics. Due to random or systematic misalignment of the slit (i.e. if the slit is not aligned with the galaxy’s major axis), there might be an offset in the measured RCs, even if the galaxy is perfectly axisymmetric. The advent of integral field unit (IFU) surveys, e.g. SAURON \citep[][]{deZeeuw2002}, ATLAS$^{\rm 3D}$ \citep[][]{Cappellari2011}, CALIFA \citep[][]{Sanchez2012}, SAMI \citep[][]{Bryant2015}, and MaNGA \citep[][]{Bundy2015}, provides two-dimensional (2D) kinematic information, reducing the effect of slit misalignment when extracting RCs \citep[e.g.][]{Yoon2021,Ristea2024a}. 

However, it was also pointed out that the RCs only work well in spirals and will underestimate the true CVCs of dispersion-dominated galaxies due to their significant non-circular motions \citep{Roper2023,Downing2023,Sands2024}. One way to overcome this intrinsic limitation of RCs and extend the studies to include early-type galaxies (ETGs) is applying corrections to account for the contribution of disordered motions, e.g. the asymmetric drift correction \citep{Shetty2020,Bershady2024} or computing circular velocities from kinetic energy \citep{Ristea2024b}. A more direct and more accurate way is to constrain the gravitational potential (or CVCs) through detailed stellar dynamical models combined with spatially resolved stellar kinematics \citep[e.g.,][]{Cappellari2013a,Li2017,Leung2018,Zhu2023}. CVCs derived from stellar dynamical models also have advantages over directly measured RCs, such as accounting for seeing effects and using accurately recovered inclination angles \citep{Cappellari2008}. 

In this paper, which is the seventh paper of our DynPop series, we derive the CVCs from the Jeans anisotropic modeling (JAM) models for 10,000 galaxies (of which 6000 are deemed reliable) provided in the MaNGA DynPop project \citep[Paper I;][]{Zhu2023}. The MaNGA DynPop project not only provides the mass distributions but also the stellar population properties \citep[Paper II;][]{Lu2023}, enabling us to study the dynamical scaling relations in conjunction with stellar population properties \citep[Paper III;][]{Zhu2024}, the density profiles for galaxy groups and clusters by combining stellar dynamics and weak lensing \citep[Paper IV;][]{Wang2024}, the variation of stellar initial mass function (IMF) in ETGs \citep[Paper V;][]{Lu2024}, and a detailed comparison of total density slopes between MaNGA and simulations \citep[Paper VI;][]{Li2024}. The structure of this paper is organized as follows. \autoref{sec:data} briefly introduces the MaNGA data and how we derive the CVCs and other galaxy properties. In \autoref{sec:CVC}, we study the correlations between CVCs (including the amplitude and the shape) and other galaxy properties (\autoref{subsec:CVCamplitude} and \autoref{subsec:CVCshape}), propose a model to quantify the shape of CVCs (\autoref{subsec:BDH_model}), and investigate the evolution of CVC shapes on the mass-size plane (\autoref{subsec:CVCshape_mass_size}). We summarize our main results in \autoref{sec:conlusions}. Throughout the paper, we assume a flat Universe with $\Omega_{\rm m} = 0.307$ and $H_0 = 67.7\,\mathrm{km~s^{-1}~Mpc^{-1}}$ \citep{Planck2016}, for consistency with the other papers of the DynPop series. In this paper, $\rm log_{10}$ is denoted by $\rm lg$ to follow the official ISO 80000-2:2019 notation.

\section{Sample and Data} \label{sec:data}
\subsection{The MaNGA survey}
As one of the three projects in Sloan Digital Sky Survey-IV \citep[SDSS-IV;][]{Blanton2017}, the Mapping Nearby Galaxies at Apache Point Observatory (MaNGA) survey \citep{Bundy2015} provides spatially resolved spectral measurements of $\sim 10,000$ nearby galaxies. The MaNGA project uses the IFU technique to obtain spectra simultaneously across the face of target galaxies, employing tightly packed fiber bundles that feed into the BOSS spectrographs \citep{Smee2013,Drory2015} on the Sloan 2.5m telescope \citep{Gunn2006}. The field of view (FoV) of MaNGA observations extends radially out to 1.5 effective radii ($R_{\rm e}$) for about two-thirds of the galaxies (Primary+ sample), and up to 2.5 $R_{\rm e}$ for roughly one-third of the galaxies (Secondary sample) at higher redshifts \citep{Law2015,Wake2017}. The final sample exhibits an approximately flat stellar mass distribution across the range $10^9-6\times10^{11} \,\rm M_{\odot}$ \citep{Wake2017}, with a median redshift of $z\sim0.03$.

The MaNGA spectra cover a wavelength range of $3600-10300\,\rm {\AA}$, with a spectral resolution of $\sigma = 72\, \rm {\rm km\,s^{-1}}$ \citep{Law2016}. The raw data are spectrophotometrically calibrated \citep{Yan2016} and processed using the Data Reduction Pipeline \citep[DRP;][]{Law2016} to produce data cubes. Stellar kinematic maps are then extracted from these data cubes through the Data Analysis Pipeline \citep[DAP;][]{Belfiore2019,Westfall2019}, which utilizes the \textsc{ppxf} software \citep{Cappellari2004,Cappellari2017,Cappellari2023} and a subset of the MILES stellar library \citep{Sanchez-Blazquez2006,Falcon-Barroso2011}, MILES-HC, to fit the absorption lines in the IFU spectra. Using a hierarchical-clustering approach, the entire MILES stellar library (985 spectra) is classified into 49 different clusters having similar spectra and therefore similar stellar parameters (e.g. effective temperature, metallicity, and surface gravity). For each cluster, all the spectra are normalized to a mean of unity and averaged without weighting to construct a representative stellar template. After excluding templates with prominent emission lines or relatively low $\rm S/N$, the final 42 stellar templates in the MILES-HC library enable reliable stellar kinematics measurements while reducing execution time by a factor of 25 \citep[see more details in section~5 of][]{Westfall2019}. Before extracting stellar kinematics, the spectra are Voronoi binned \citep{Cappellari2003} to a signal-to-noise ratio of $\rm S/N=10$ to obtain reliable measurements.

\subsection{Dynamical models, CVCs, and other dynamical properties}\label{subsec:Dyn}

We performed JAM \citep[][]{Cappellari2008,Cappellari2020} to construct dynamical models for the whole MaNGA sample in \citet{Zhu2023}. The JAM model allows for anisotropy in second velocity moments and two different assumptions on the orientation of the velocity ellipsoid, i.e. $\rm JAM_{cyl}$ (cylindrically-aligned) and $\rm JAM_{sph}$ (spherically-aligned). Four different mass models are adopted in \citet{Zhu2023}, which differ primarily in their assumptions about DM distributions: (i) the mass-follows-light model which assumes that the total mass density traces the luminosity density (hereafter MFL model), (ii) the model which assumes a spherical Navarro-Frenk-White \citep[NFW;][]{Navarro1996} dark halo (hereafter NFW model), (iii) the fixed NFW model which assumes a spherical NFW halo predicted by the stellar mass-to-halo mass relation in \citet{Moster2013} and mass-concentration relation in \citet{Dutton2014} (hereafter fixed NFW model), and (iv) the model which assumes a generalized NFW \citep{Wyithe2001} dark halo (hereafter gNFW model). Further details about the eight models can be found in \citet{Zhu2023}, while the data catalog of dynamical properties is available online.\footnote{As supplementary files of \citet{Zhu2023} on the journal website and on the website of MaNGA DynPop at \url{https://manga-dynpop.github.io/}}

Based on comparisons of observed and modeled stellar kinematics, the entire sample is classified according to different modeling qualities (Qual = $-1$, 0, 1, 2, 3 from worst to best). In this work, we select 6065 galaxies that are flagged as Qual $ \geqslant 1$, for which the dynamical quantities related to the total mass distribution are nearly insensitive to variations in model assumptions \citep{Zhu2023}. Throughout this paper, we adopt the gNFW model, which is the most flexible mass model, with the $\rm JAM_{cyl}$ assumption (the results and conclusions remain consistent when adopting the $\rm JAM_{sph}$ assumption) unless stated otherwise. 

The total mass distribution consists of three components: the nuclear supermassive black hole, the stellar mass distribution, and the DM mass distribution (i.e. a gNFW dark halo in this model). The black hole mass (assuming to be a point mass) is estimated from $M_{\rm BH}-\sigma_{\rm c}$ relation \citep{McConnell2011}, where $\sigma_{\rm c}$ is computed as the mean stellar velocity dispersion within 1 FWHM of the MaNGA point-spread function (PSF). For the stellar component, we use the Multi-Gaussian Expansion \citep[MGE;][]{Emsellem1994,Cappellari2002} method to fit SDSS $r$-band images and obtain the surface brightness. Then the surface brightness is deprojected to obtain the luminosity density of the kinematic tracer in the three-dimensional space and further the stellar mass distributions when multiplied by the stellar mass-to-light ratio. We note that the MGE is a special case of the Gaussian mixture model \citep{Fraley2002}, where the Gaussians all have the same center and, in our case, the same orientations. The typical number of Gaussian components in our MGE models is 15, which has been well tested to be good enough (with 1$\sigma$ error of 10\%) for galaxy photometry and  dynamical modeling \citep[][section~4.1.1]{Cappellari2013a}. Examples of MGE models for the stellar component have been presented in \citet[][figure~2]{Zhu2023}. Following \citet{Cappellari2013a}, the gNFW profile is written as
\begin{equation}
    \rho_{_{\rm DM}}(r) = \rho_s\left(\frac{r}{r_s}\right)^{\gamma}\left(\frac{1}{2}+\frac{1}{2}\frac{r}{r_s}\right)^{-\gamma-3},
\end{equation}
where $r_s$ is the characteristic radius, $\rho_s$ is the characteristic density, and $\gamma$ is the inner density slope. This profile follows the same functional form as the commonly used gNFW profile \citep[e.g. equation 1 in][]{Wyithe2001}, while its amplitude linearly scales by a factor of $\left(\frac{1}{2}\right)^{-\gamma-3}$ for a given $\gamma$ ($\rho_s$ and $\gamma$ are independent parameters).

\begin{figure*}
    \centering
    \includegraphics[width=1\textwidth]{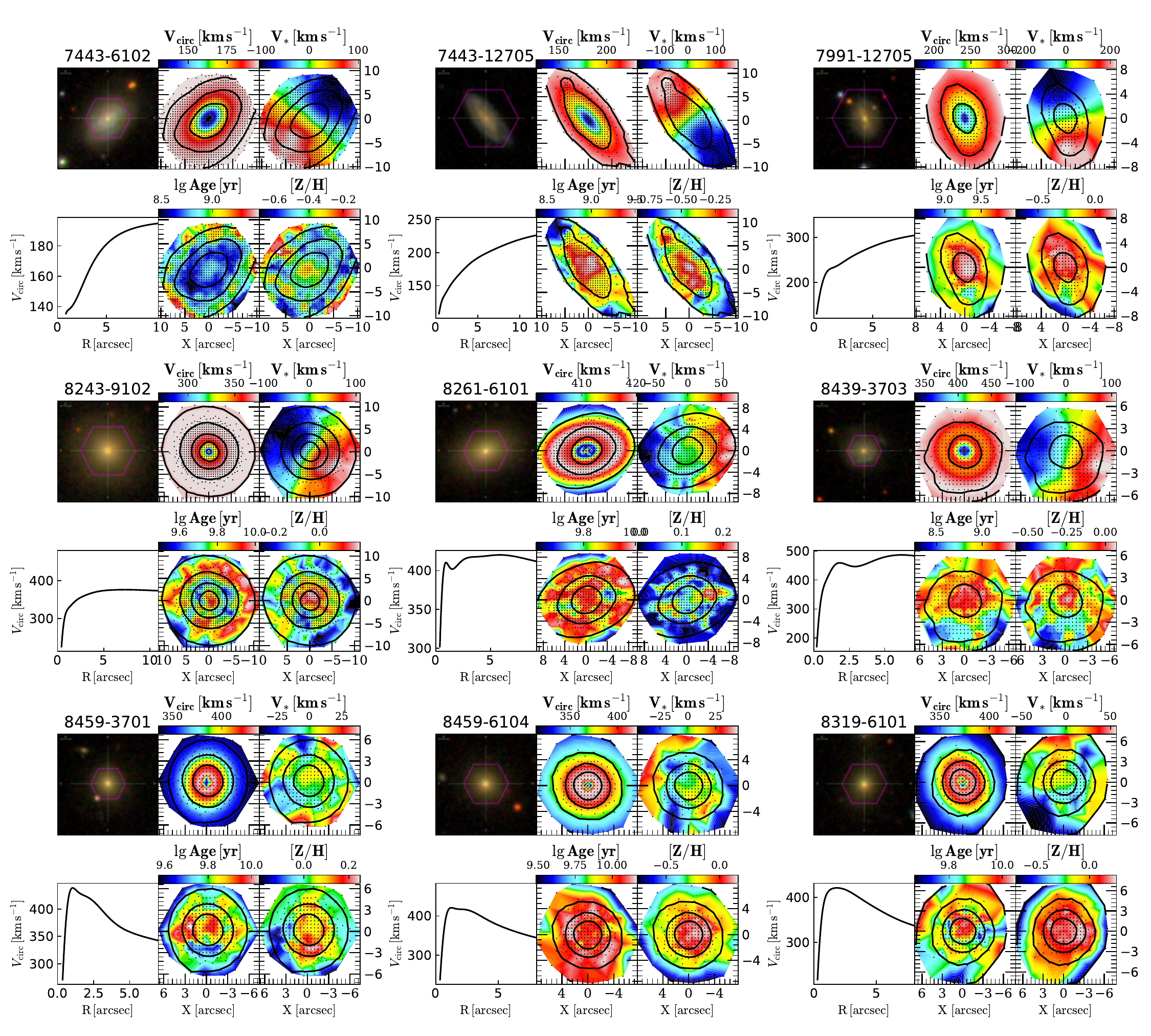}
    \caption{Examples of different shapes of JAM-derived CVCs (rising CVCs for the first two rows, flat CVCs for the second two rows, and declining CVCs for the third two rows). For each galaxy, there are six images: the RGB image with plateifu (top left), the map of circular velocities (top middle), the map of line-of-sight velocity (top right), the CVC in the equatorial plane derived from JAM model (bottom left), the map of stellar age (bottom middle), the map of stellar metallicity (bottom right). The black contours are the observed surface brightness contours in steps of 1 mag. The black dots are the centroids of the Voronoi bins from which the maps were linearly interpolated. }
    \label{fig:example_maps}
\end{figure*}

With the JAM-determined gravitational potential $\Phi(R, z)$ expressed in the form of MGE, we use the \textsc{mge\_vcirc} procedure \citep[equation 45 in ][]{Cappellari2020} in the \textsc{jampy} package to calculate the circular velocity $V_{\rm circ}$ at a given galactocentric radius $R$ in the equatorial plane. Assuming that the galaxies are axisymmetric, we further derive the maps with constant values of $V_{\rm circ}$  at a given galactocentric radius $R=\sqrt{x_{\rm bin}^2 + y_{\rm bin}^2/q^{2}}$, where ($x_{\rm bin}$, $y_{\rm bin}$) are the coordinates of Voronoi bins and $q\equiv b/a$ is the axial ratio of a half-light elliptical isophote derived from MGE. In \autoref{fig:example_maps}, we present examples of CVCs, circular velocity maps, line-of-sight velocity maps, stellar age maps and stellar metallicity maps for galaxies with different CVC shapes (rising, flat, or declining in the outskirts). \autoref{tab:CVC} presents the parameters of CVCs, including the circular velocity at half-light radius and the maximum circular velocity within kinematic data range. In \autoref{tab:CVC_error}, we provide the systematic uncertainties arising from model differences, following the calculation presented in table~3 of \citet{Zhu2023}.

\begin{deluxetable*}{llcccccccccccc}
\tabletypesize{\scriptsize}
\tablewidth{0pt} 
\tablecaption{Parameters of circular velocity curves for 6000 nearby galaxies. \label{tab:CVC}}
\tablehead{
\colhead{plateifu} & \colhead{mangaid}& \colhead{DA} & \colhead{$R_{\rm e}$} & \colhead{$R_{\rm e}^{\rm maj}$} & \colhead{$\rm rFWHM\_IFU$} & \colhead{$R(V_{\rm circ}^{\rm max})$} & \colhead{rmax} & \colhead{$V_{\rm circ}(R_{\rm e})$} & \colhead{$V_{\rm circ}(R_{\rm e}^{\rm maj})$} & \colhead{$V_{\rm circ}^{\rm max}$} & \colhead{$V_{\rm circ}({\rm rmax})$} & \colhead{$\lg\,M_{\rm BH}$} & \colhead{Qual}\\
\colhead{} & \colhead{} & \colhead{(Mpc)} & \colhead{(arcsec)} & \colhead{(arcsec)} & \colhead{(arcsec)} & \colhead{(arcsec)} & \colhead{(arcsec)} & \colhead{($\rm
km~s^{-1}$)} & \colhead{($\rm km~s^{-1}$)} & \colhead{($\rm km~s^{-1}$)} & \colhead{($\rm km~s^{-1}$)} & \colhead{($\rm M_{\odot}$)}& \colhead{}
}
\colnumbers
\startdata 
        7443-1901  & 12-84620   & 81.139  & 4.291  & 4.789  & 2.580  & 5.750  & 5.750  & 108.584 & 116.479 & 130.848 & 130.848 & 4.580 & 0 \\
        7443-6103  & 12-84665   & 79.374  & 5.769  & 8.737  & 2.557  & 10.332 & 10.332 & 112.148 & 141.332 & 152.934 & 152.934 & 5.814 & 0 \\
        7443-3702  & 12-84670   & 428.710 & 3.295  & 3.298  & 2.540  & 2.674  & 6.456  & 404.585 & 404.550 & 411.914 & 377.347 & 8.389 & 2 \\
        7443-1902  & 12-49536   & 81.841  & 3.566  & 4.766  & 2.573  & 5.885  & 5.885  & 91.468  & 97.980  & 102.546 & 102.546 & 5.217 & 0 \\
        7443-9101  & 12-84660   & 170.514 & 6.951  & 7.461  & 2.591  & 11.591 & 11.591 & 109.817 & 110.667 & 121.424 & 121.424 & 5.911 & 0 \\
        7443-12702 & 12-84674   & 237.488 & 9.070  & 9.253  & 2.561  & 9.504  & 9.504  & 111.012 & 111.123 & 111.241 & 111.241 & 5.758 & 0 \\
        7443-12704 & 12-84731   & 81.049  & 13.045 & 26.411 & 2.537  & 3.334  & 14.369 & 131.025 & 159.826 & 141.175 & 133.302 & 6.388 & 3 \\
        7443-6102  & 12-180432  & 119.823 & 7.328  & 8.898  & 2.558  & 10.315 & 10.315 & 190.581 & 193.384 & 194.786 & 194.786 & 6.635 & 3 \\
        7443-3701  & 12-193534  & 77.483  & 4.717  & 6.054  & 2.596  & 6.285  & 7.374  & 80.656  & 82.244  & 82.247  & 81.593  & 5.309 & 0 \\
        7443-12701 & 12-98126   & 88.386  & 4.756  & 5.825  & 2.581  & 9.437  & 9.437  & 118.948 & 119.206 & 120.597 & 120.597 & 5.810 & 1 \\
\enddata
\tablecomments{The dynamical quantities in this table are derived from $\rm JAM_{cyl}$ + gNFW model (see \autoref{subsec:Dyn}). (1) plateifu: The name of plate ID + IFU design ID (e.g. 7443-1901); (2) mangaid: unique MaNGA ID (e.g. 12-84620); (3) DA: angular diameter distance that assumes a flat Universe with $\Omega_{\rm m}=0.307$ and $h=0.677$ \citep{Planck2016}; (4) $R_{\rm e}$: effective radius (projected circular half-light radius from MGE fitting, in SDSS $r$ band); (5) $R_{\rm e}^{\rm maj}$: major axis of elliptical half-light isophote from MGE fitting, in SDSS $r$ band; (6) $\rm rFWHM\_IFU$: the PSF FWHM values of IFU observations, in SDSS $r$ band; (7) $R(V_{\rm circ}^{\rm max})$: the galactocentric radius where CVC has the maximum circular velocity; (8) rmax: the kinematic data range, which is defined as the largest radius of the Voronoi bins; (9) $V_{\rm circ}(R_{\rm e})$: circular velocity at half-light radius in the equatorial plane; (10) $V_{\rm circ}^{\rm maj}$: circular velocity at the major axis of half-light ellipse in the equatorial plane; (11) $V_{\rm circ}^{\rm max}$: the maximum circular velocity within kinematic data range; (12) $V_{\rm circ}({\rm rmax})$: the circular velocity at rmax; (13) $\lg\,M_{\rm BH}$: the mass of central supermassive black hole adopted in our dynamical models, which is derived from scaling relation (see Section~\ref{subsec:Dyn}); (14) Qual: the quality of JAM models ($-1$ to 3 from worst to best), only galaxies with Qual $\geqslant1$ have reliable CVC measurements.\\
The full catalog is available as a machine-readable table.}
\end{deluxetable*}
\begin{deluxetable*}{cccccccccccccc}
\tabletypesize{\small}
\tablewidth{0pt}
\tablecaption{Systematic errors of circular velocity measurements \label{tab:CVC_error}}
\tablehead{
\colhead{Quantities} & \multicolumn{3}{c}{Qual = 0} & \multicolumn{3}{c}{Qual = 1} & \multicolumn{3}{c}{Qual = 2} & \multicolumn{3}{c}{Qual = 3} & \colhead{Mass Models} \\
\colhead{} & \colhead{Slope} & \colhead{$\Delta$} & \colhead{Error} & \colhead{Slope} & \colhead{$\Delta$} & \colhead{Error} & \colhead{Slope} & \colhead{$\Delta$} & \colhead{Error} & \colhead{Slope} & \colhead{$\Delta$} & \colhead{Error} & \colhead{} \\
\colhead{} & \colhead{} & \colhead{(dex)} & \colhead{} & \colhead{} & \colhead{(dex)} & \colhead{} & \colhead{} & \colhead{(dex)} & \colhead{} & \colhead{} & \colhead{(dex)} & \colhead{} & \colhead{}
}
\colnumbers
\startdata
$V_{\rm circ}(R_{\rm e})$      & 1.00    & 0.022  & 3.65\%  & 1.00    & 0.0098 & 1.61\%  & 1.00    & 0.018  & 2.97\%  & 1.00    & 0.0047 & 0.77\%  & NFW, gNFW \\
$V_{\rm circ}(R_{\rm e}^{\rm maj})$ & 1.00  & 0.020   & 3.31\%  & 1.00  & 0.0099 & 1.62\%  & 1.01  & 0.018  & 2.97\%  & 1.00  & 0.0060 & 0.98\%  & NFW, gNFW \\
$V_{\rm circ}^{\rm max}$       & 1.00  & 0.022  & 3.65\%  & 1.00  & 0.011  & 1.81\%  & 1.01  & 0.018  & 2.97\%  & 1.00  & 0.0056 & 0.92\%  & NFW, gNFW \\
$V_{\rm circ}({\rm rmax})$     & 1.01  & 0.024  & 3.98\%  & 1.00  & 0.011  & 1.81\%  & 1.01  & 0.018  & 2.97\%  & 1.00  & 0.0061 & 1.00\%  & NFW, gNFW \\
\enddata
\tablecomments{The systematic errors of circular velocities ($V_{\rm circ}(R_{\rm e})$, $V_{\rm circ}(R_{\rm e}^{\rm maj})$, $V_{\rm circ}^{\rm max}$, $V_{\rm circ}({\rm rmax})$) for different quality groups, following the calculation presented in table~3 of \citet{Zhu2023}. The slope and $\Delta$ denote the slope and observed scatter obtained by the \textsc{lts\_linefit} procedure (with \texttt{clip=6}). The error is defined as $\Delta/\sqrt{2}$ assuming that the quantities on both axes are comparable. The errors of quantities in this table are derived from four models (i.e. the more flexible NFW and gNFW models, each with two orientations of the velocity ellipsoid).}
\end{deluxetable*}

Other structural and dynamical quantities used in this work are mainly taken from \citet{Zhu2023}, which are derived directly from observational data and the best-fitting dynamical models (with five free parameters for the most flexible gNFW model). We briefly introduce the parameters and the corresponding keywords here: 
\begin{itemize}
    \item Ellipticity $\varepsilon$ (keyword: \texttt{Eps\_MGE}): ellipticity of the half-light elliptical isophote derived from the MGE model.
    \item Size parameters $R_{\rm e}$ (keyword: \texttt{Re\_arcsec\_MGE}) and $R_{\rm e}^{\rm maj}$ (keyword: \texttt{Rmaj\_arcsec\_MGE}): $R_{\rm e}$ is the circularized half-light radius (effective radius), while $R_{\rm e}^{\rm maj}$ is the semi-major axis of the half-light elliptical isophote. Both $R_{\rm e}$ and $R_{\rm e}^{\rm maj}$ in the catalog of \citet{Zhu2023} have been scaled by a factor of 1.35 following \citet{Cappellari2013a}.
    \item Total luminosity in SDSS $r$-band $L$ (keyword: \texttt{Lum\_tot\_MGE}): $L$ is derived from the MGE model of the SDSS $r$-band image and has been corrected for the dust extinction \citep[see more details about dust extinction in][]{Lu2023}.
    \item Effective stellar velocity dispersion $\sigma_{\rm e}$ (keyword: \texttt{Sigma\_Re}): the second moment of the line-of-sight velocity within the elliptical half-light isophote (with an area of $\pi R_{\rm e}^2$), defined as 
    \begin{equation}\label{eq:sigmae}
    \sigma_{\rm e} \equiv \langle V_{\rm rms}^2\rangle_{\rm e}^{1/2} \approx \sqrt{\frac{\sum_k F_k (V_k^2+\sigma_k^2)}{\sum_k F_k}},
    \end{equation}
    where $F_k$, $V_k$, and $\sigma_k$ are the flux, stellar velocity, and stellar velocity dispersion in the $k$th IFU spaxel.
    This quantity closely approximates the velocity dispersion $\sigma$ that one would measure by coadding all spectra within the same aperture, and fitting $(V,\sigma)$ for that spectrum \citep[section~4.3]{Cappellari2013a}.
    
    \item Stellar angular momentum proxy or spin parameter $\lambda_{\rm R_e}$ (keyword: \texttt{Lambda\_Re}): $\lambda_{\rm R_e}$ is defined within the same aperture as $\sigma_{\rm e}$ (i.e. the elliptical half-light isophote), written as \citep{Emsellem2007}
    \begin{equation}\label{eq:lambdaRe}
    \lambda_{R_{\rm e}} = \frac{\sum_k F_k R_k|V_k|}{\sum_k F_k R_k \sqrt{V_k^2+\sigma_k^2}},
    \end{equation}
where $F_k$, $V_k$ and $\sigma_k$ are the same as \autoref{eq:sigmae}; $R_k$ is the distance of $k$th spaxel to the galaxy center. The $\lambda_{\rm R_e}$ in the catalog of \citet{Zhu2023} has been corrected for the beam-smearing effect following \citet{Graham2018}\footnote{\url{https://github.com/marktgraham/lambdaR_e_calc}}.
    \item Dynamical mass $M_{\rm JAM}$: $M_{\rm JAM}$ is defined as 
    \begin{equation}
    \label{eq:Mjam}
    M_{\rm JAM} \equiv (M/L)_{\rm JAM}\times L,
    \end{equation}
where $(M/L)_{\rm JAM}$ is the dynamical mass-to-light ratio (keyword: \texttt{log\_ML\_dyn}) from the $\rm JAM_{cyl}$+MFL model and $L$ is the SDSS $r$-band total luminosity (keyword: \texttt{Lum\_tot\_MGE}).
    \item Morphology and photometric properties T-Type, $B/T$, $n_{\rm Ser, bulge}$, $R_{\rm e,bulge}/R_{\rm e, disk}$: these parameters (in the SDSS $r$-band) are drawn from the PyMorph photometric and deep-learning morphological catalogs \citep{Fischer2019,Dominguez-Sanchez2022}. The T-Type values (keyword: \texttt{TType}), ranging from $-4$ to 9, correspond to ETGs through to late-type galaxies (LTGs). The bulge-to-total luminosity ratio $B/T$ (keyword: \texttt{BT\_SE}), the Sersic index \citep{Sersic1968} of the bulge component $n_{\rm Ser, bulge}$ (keyword: \texttt{N\_SE\_BULGE}), and the ratio of the effective radii between the bulge and disk components ($R_{\rm e, bulge}/R_{\rm e, disk}$) are derived from two-component (Sersic + exponential) fits. Here, $R_{\rm e, bulge}$ is the circularized effective radius of the bulge component calculated by $\sqrt{\texttt{A\_hl\_SE\_BULGE}^2\times\texttt{BA\_SE\_BULGE}}$, where \texttt{A\_hl\_SE\_BULGE} is the bulge half-light semi-major axis and \texttt{BA\_SE\_BULGE} is the bulge axis ratio (semi-minor/semi-major). Similarly, $R_{\rm e, disk}$ is the circularized effective radius of the disk component calculated by $\sqrt{\texttt{A\_hl\_SE\_DISK}^2\times\texttt{BA\_SE\_DISK}}$, where \texttt{A\_hl\_SE\_DISK} is the disk half-light semi-major axis and \texttt{BA\_SE\_DISK} is the disk axis ratio (semi-minor/semi-major). We adopt the criteria \texttt{FLAG\_FIT}$\neq$3
    and \texttt{FLAG\_FAILED\_SE}=0 to exclude galaxies with bad/failed two-component fits and remove the flipped galaxies with \texttt{N\_SE\_BULGE}=1 and \texttt{N\_SE\_DISK}$\leq$1 \citep[see details in section~2.1.3 of][]{Fischer2019}. Regarding $n_{\rm Ser, bulge}$, we further discard galaxies with \texttt{N\_SE\_BULGE}=8 and \texttt{BT\_SE}$\leq$0.1, as the former are likely failed fits that hit the boundary of bulge Sersic index, while the latter may have a bulge fraction too small to yield a reliable measurement of the bulge Sersic index.
    \item The ratio between the scale radius of the NFW halo and the luminous half-light radius $r_{\rm s}/R_{\rm e}$: $r_{\rm s}$ is estimated from the Chabrier IMF-based stellar mass, which is converted from the Salpeter IMF-based stellar mass $M_{\rm \ast}$ (see the definition in \autoref{subsec:SPS}) by subtracting 0.215 dex \citep[][figure~4]{Madau2014}, the stellar mass-to-halo mass relation in \citet{Moster2013}, and the mass-concentration relation in \citet{Dutton2014}.
\end{itemize}

\subsection{Stellar population properties based on Stellar Population Synthesis (SPS)}\label{subsec:SPS}
The stellar population properties used in this work come from \citet{Lu2023}. They fit the IFU spectra of the MaNGA DRP \citep{Law2016} data cubes using the \textsc{ppxf} software \citep{Cappellari2004,Cappellari2017,Cappellari2023} with the \textsc{fsps} models \citep{Conroy2009,Conroy2010}, the Salpeter \citep{Salpeter1955} IMF, and the MIST isochrones \citep{Choi2016}. The properties are briefly summarized below:
\begin{itemize}
    \item Luminosity-weighted stellar age $\langle\lg \rm Age\rangle$ (keyword: \texttt{LW\_Age\_Re}), luminosity-weighted stellar metallicity $\langle[Z/H]\rangle$ (keyword: \texttt{LW\_Metal\_Re}) and their radial gradients (keywords: \texttt{LW\_Age\_Slope} and \texttt{LW\_Metal\_Slope}): the luminosity-weighted values are calculated as
    \begin{equation}
    \langle x\rangle \equiv \frac{\sum^{N}_{i=1}w_i L_{i}x_{i}}{\sum^{N}_{i=1}w_i L_{i}},
    \end{equation}
    where $w_{i}$ is the fraction of mass contributed by the $i$th template (the \textsc{ppxf} fit weight, if the templates are normalized to unitary mass), $L_{i}$ is the SDSS r-band luminosity per unit mass of the $i$th template, and $x_{i}$ is the $\rm \lg Age$ (or $\mathrm{[Z/H]}$) of the $i$th template. The radial gradients are measured by linearly fitting the $\langle\lg \rm Age\rangle$ or $\langle[Z/H]\rangle$ profiles within the elliptical half-light isophote.
    
    \item Star formation histories $T_{50}$ (keyword: \texttt{T50}) and $T_{90}$ (keyword: \texttt{T90}): $T_{50}$ and $T_{90}$ are defined as the cosmic times when the stars that account for 50\% and 90\% of galaxies' present-day stellar mass are formed, respectively. Note that $T_{50}$ and $T_{90}$ are given as lookback times in the catalog of \citet{Lu2023} and here we convert them into cosmic time based on the cosmological parameters used in this work.
    
    \item Averaged intrinsic stellar mass-to-light ratio $(M_{\ast}/L)_{\rm SPS}$ (keyword: \texttt{ML\_int\_Re}) within the elliptical half-light isophote, which is calculated as 
    \begin{equation}
    (M_{\ast}/L)_{\rm SPS}=\frac{\sum_{i=1}^{N} w_{i} M_{i}^{\rm nogas}}{\sum_{i=1}^{N} w_{i} L_{i}},
    \end{equation}
    where $M_{i}^{\rm nogas}$ is the stellar mass of the $i$th template, which includes the mass of living stars and stellar remnants but excludes the mass of lost gas during stellar evolution.
    
    \item Total stellar mass $M_{\ast}$ defined as 
    \begin{equation}
    \label{eq:MsSPS}
    M_{\ast} = (M_{\ast}/L)_{\rm SPS} \times L,
    \end{equation}
    where $(M_{\ast}/L)_{\rm SPS}$ is the SDSS $r$-band stellar mass-to-light ratio derived from the stacked spectrum within the elliptical half-light isophote (keyword: \texttt{ML\_int\_Re}), and $L$ is the total luminosity derived from MGE model (keyword: \texttt{Lum\_tot\_MGE}). 

\end{itemize}

\section{The circular velocity curves across various galaxy types}\label{sec:CVC}

Based on the JAM-derived CVCs of 238 galaxies in the CALIFA survey \citep{Sanchez2012}, \citet{Kalinova2017} studied the amplitude and the shape of CVCs and their correlations with galaxy properties (e.g. mass, luminosity, morphology, stellar age, and stellar metallicity). Recently, \citet{Ristea2024b} computed the circular velocities (both stellar and gas) for a subset of MaNGA galaxies ($\sim$3500) by accounting for the contribution from disordered motions to the kinetic energy. For different stellar mass bins, they derived the empirical relations between the rotation velocities at 1.3$R_{\rm e}$ and luminosity-weighted rotational-to-dispersion velocity ratios $V_{\rm rot}/\sigma$ \citep[][equation~10]{Cappellari2007} within the same aperture, and extrapolated them to $V_{\rm rot}/\sigma=10$ to obtain corresponding ‘asymptotic’ rotational velocities. Using the data products of the MaNGA DynPop project \citep{Zhu2023,Lu2023}, which significantly increases the sample size of stellar dynamical models to 10,000 galaxies for the first time, we are able to directly derive CVCs and gain a comprehensive understanding of the correlations between the inner gravitational potential (reflected by the amplitude and shape of CVCs) and other galaxy properties in a statistical way. 

\subsection{The amplitude of CVCs}\label{subsec:CVCamplitude}
The amplitude of CVCs is usually characterized by the maximum (asymptotic) circular velocity \citep{Tully1977} for spirals, while the characteristic velocity measurement for ETGs is typically the velocity dispersion which enters the Faber-Jackson relation \citep{Faber1976} and fundamental plane \citep{Djorgovski1987,Dressler1987}. Previous studies aimed at unifying the dynamical relations of spirals and ETGs were done by converting the velocity dispersion $\sigma_{\rm e}$ into the circular velocity at half-light radius $V_{\rm circ}(R_{\rm e}^{\rm maj})$ or the maximum circular velocity $V_{\rm circ}^{\rm max}$ \citep{Padmanabhan2004,Courteau2007,Schulz2010,Dutton2011,Cappellari2013a}. Here, we revisit the empirical relations (in the form $V_{\rm circ}=k \sigma_{\rm e}$, where $k$ is the conversion factor) using the accurate $\sigma_{\rm e}$ provided by MaNGA \citep{Law2021}, together with the circular velocities derived from our dynamical models. To avoid potential overestimation of circular velocities (or total masses) for nearly face-on galaxies \citep[e.g. fig.~12 in ][]{Lablanche2012}, we further exclude those galaxies with inclination angles below 30$^\circ$ (accounting for $\sim 15\%$ of galaxies with Qual $ \geqslant 1$) in this section. This exclusion has a negligible impact on the best-fitting relations, altering the coefficients and scatter by only a few percent.

\begin{figure*}
    \centering
    \includegraphics[width=0.8\textwidth]{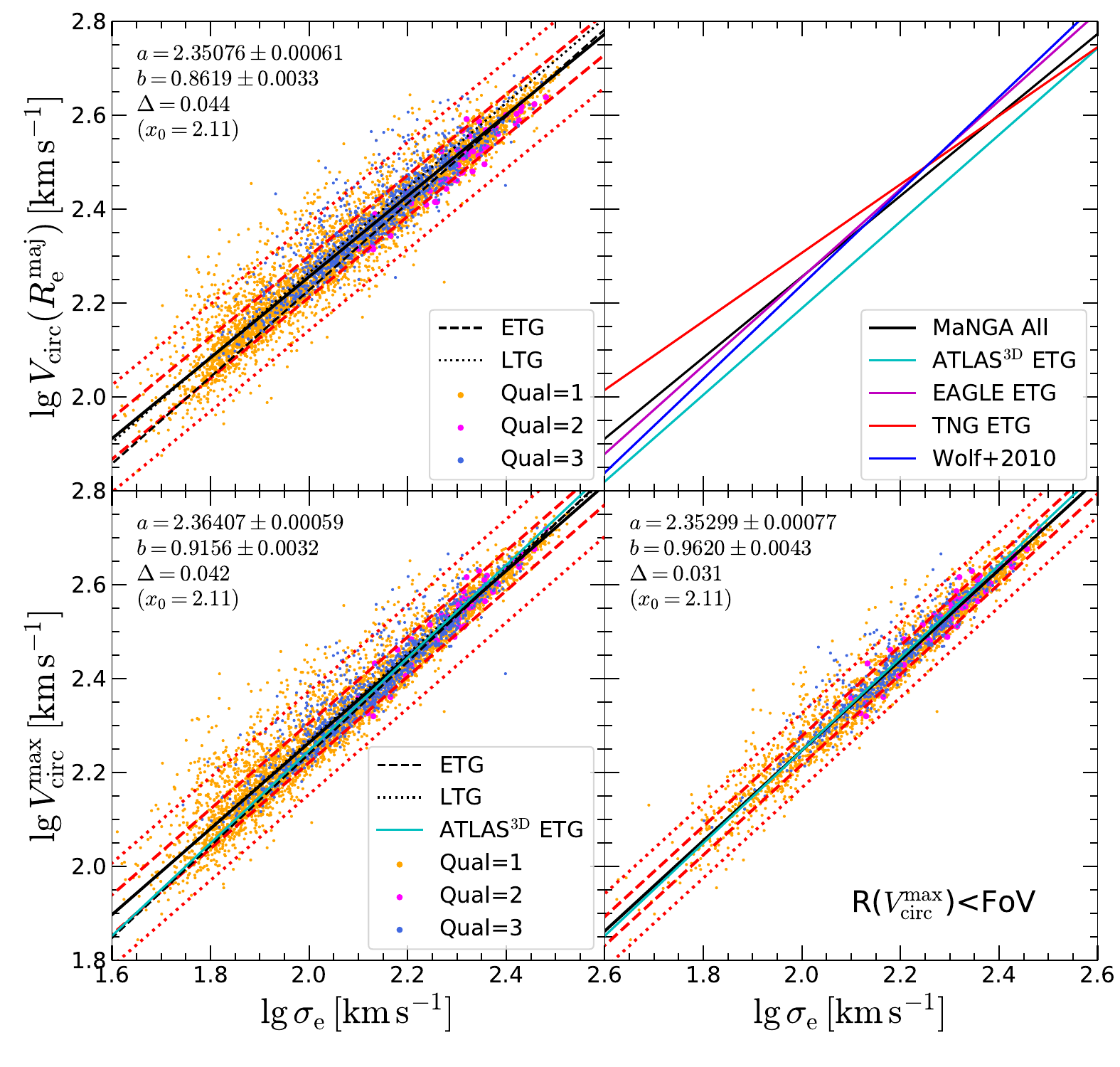}
    \caption{The correlations between $\sigma_{\rm e}$ and circular velocity at the half-light radius $V_{\rm circ}(R_{\rm e}^{\rm maj})$ (top panels) as well as the maximum circular velocity $V_{\rm circ}^{\rm max}$ (bottom panels). These relations are derived from $\rm Qual\geqslant1$ galaxies with nearly face-on galaxies (inclination angles below 30$^\circ$) excluded. \textit{Top left}: The relation between $\sigma_{\rm e}$ and $V_{\rm circ}(R_{\rm e}^{\rm maj})$. Symbols with different color correspond to different JAM model qualities \citep[see section~5.1 in][]{Zhu2023}. The coefficients of the best-fitting line $\rm y=a+b(x-x_0)$ and the observed rms scatter $\Delta$ are obtained from the \textsc{lts\_linefit} procedure (with \texttt{clip=4}) \citep{Cappellari2013a}. The black solid, red dashed, and red dotted lines represent the best-fitting, the 1$\sigma$ error (68\%), and the 2.6$\sigma$ error (99\%), respectively. The best-fitting relation can be approximated as $V_{\rm circ}(R_{\rm e}^{\rm maj})\approx1.62\times\sigma_{\rm e}$ (\autoref{eq:Vcirc_Rmaj}) with an error of $\Delta/\sqrt{2}=7$\%. The black dashed and black dotted lines correspond to the best-fitting relations for ETGs ($k\approx1.59$) and LTGs ($k\approx1.71$), respectively. \textit{Top right}: Comparisons with ATLAS$^{\rm 3D}$ (cyan line), EAGLE simulation (magenta line), Illustris-TNG simulation (red line), and theoretical prediction of \citet[][black dashed line]{Wolf2010}, which are taken from \citet{Ferrero2021}. \textit{Bottom left}: Similar to the top-left panel but showing the correlation between $\sigma_{\rm e}$ and $V_{\rm circ}^{\rm max}$. The best-fitting relation can be described as $V_{\rm circ}^{\rm max}\approx1.72\times\sigma_{\rm e}$ with an error of $\Delta/\sqrt{2}=7$\%. The black dashed and black dotted lines correspond to the best-fitting relations for ETGs and LTGs, respectively. \textit{Bottom right}: Similar to the bottom-left panel but only including the galaxies whose radius of maximum circular velocity lies in the kinematic data range. The best-fitting relation remains nearly unchanged with a slightly smaller error of 5\%.}
    \label{fig:VcRmaj_Vmax_sigma}
\end{figure*}

The correlation between $\sigma_{\rm e}$ and $V_{\rm circ}(R_{\rm e}^{\rm maj})$ in MaNGA is presented in the top-left panel of \autoref{fig:VcRmaj_Vmax_sigma}. The best-fitting relation is given by
\begin{equation}
    \lg V_{\rm circ}(R_{\rm e}^{\rm maj})=2.35+0.86\times(\lg\sigma_{\rm e}-2.11),
\end{equation}
or approximately equal to
\begin{equation}\label{eq:Vcirc_Rmaj}
    V_{\rm circ}(R_{\rm e}^{\rm maj})\approx3.43\times\sigma_{\rm e}^{0.86}\approx1.62\times\sigma_{\rm e},
\end{equation}
with an error of $\Delta/\sqrt{2}=7$\%. When classifying the sample into different morphological types following \citet[][section~2.5]{Zhu2024}, the conversion factors for ETGs ($k\approx1.59$) and LTGs ($k\approx1.71$) are similar. Comparisons with previous observation \citep[ATLAS$^{\rm 3D}$;][]{Cappellari2013a}, the EAGLE \citep{Schaye2015,Crain2015} and Illustris-TNG \citep{Naiman2018,Springel2018,Pillepich2018,Marinacci2018,Nelson2018} cosmological simulations \citep[the relations taken from][]{Ferrero2021}, and theoretical prediction \citep{Wolf2010} are shown in the top-right panel. \citet{Cappellari2013a} found a slightly smaller factor $k\approx1.51$ in ETGs, while \citet{Wolf2010} predicted $k=\sqrt{3}$. The relation in the EAGLE simulations also has a similar factor of $k\approx 1.72$ (or equivalently $V_{\rm circ}(R_{\rm e})=2.37\times\sigma_{\rm e}^{0.94}$). However, unlike the relations mentioned above, which show weak $\sigma_{\rm e}$ dependence, the relation in Illustris-TNG simulations ($V_{\rm circ}(R_{\rm e})=7.03\times\sigma_{\rm e}^{0.73}$) has an exponent significantly deviating away from 1.

A slightly tighter correlation between $\sigma_{\rm e}$ and $V_{\rm circ}^{\rm max}$ is observed in the bottom-left panel of \autoref{fig:VcRmaj_Vmax_sigma}, with
\begin{equation}
    \lg V_{\rm circ}^{\rm max}=2.36+0.92\times(\lg\sigma_{\rm e}-2.11),
\end{equation}
or approximately
\begin{equation}
    V_{\rm circ}^{\rm max}\approx2.62\times\sigma_{\rm e}^{0.92}\approx1.72\times\sigma_{\rm e},
\end{equation}
with an error of $\Delta/\sqrt{2}=7$\%. No significant difference in the conversion factors are found for ETGs ($k\approx1.71$) and LTGs ($k\approx1.75$), as shown in the bottom-left panel of \autoref{fig:VcRmaj_Vmax_sigma}. This result is consistent with the conversion factor found for ETGs in ATLAS$^{\rm 3D}$ \citep[$k\approx1.76$;][]{Cappellari2013a}. Note that $V_{\rm circ}^{\rm max}$ is defined as the maximum circular velocity within the region where we have stellar kinematic data (usually 1.5 $R_{\rm e}$ or 2.5 $R_{\rm e}$ in MaNGA). When we only consider the galaxies whose radius of maximum circular velocity lies in the kinematic data range, the conversion factor $k\approx1.72$ remains nearly unchanged but with a smaller error of $\Delta/\sqrt{2}=5$\% (bottom-right panel of \autoref{fig:VcRmaj_Vmax_sigma}).

\subsection{The shape of CVCs}\label{subsec:CVCshape}

\begin{figure*}
    \centering
    \includegraphics[width=\textwidth]{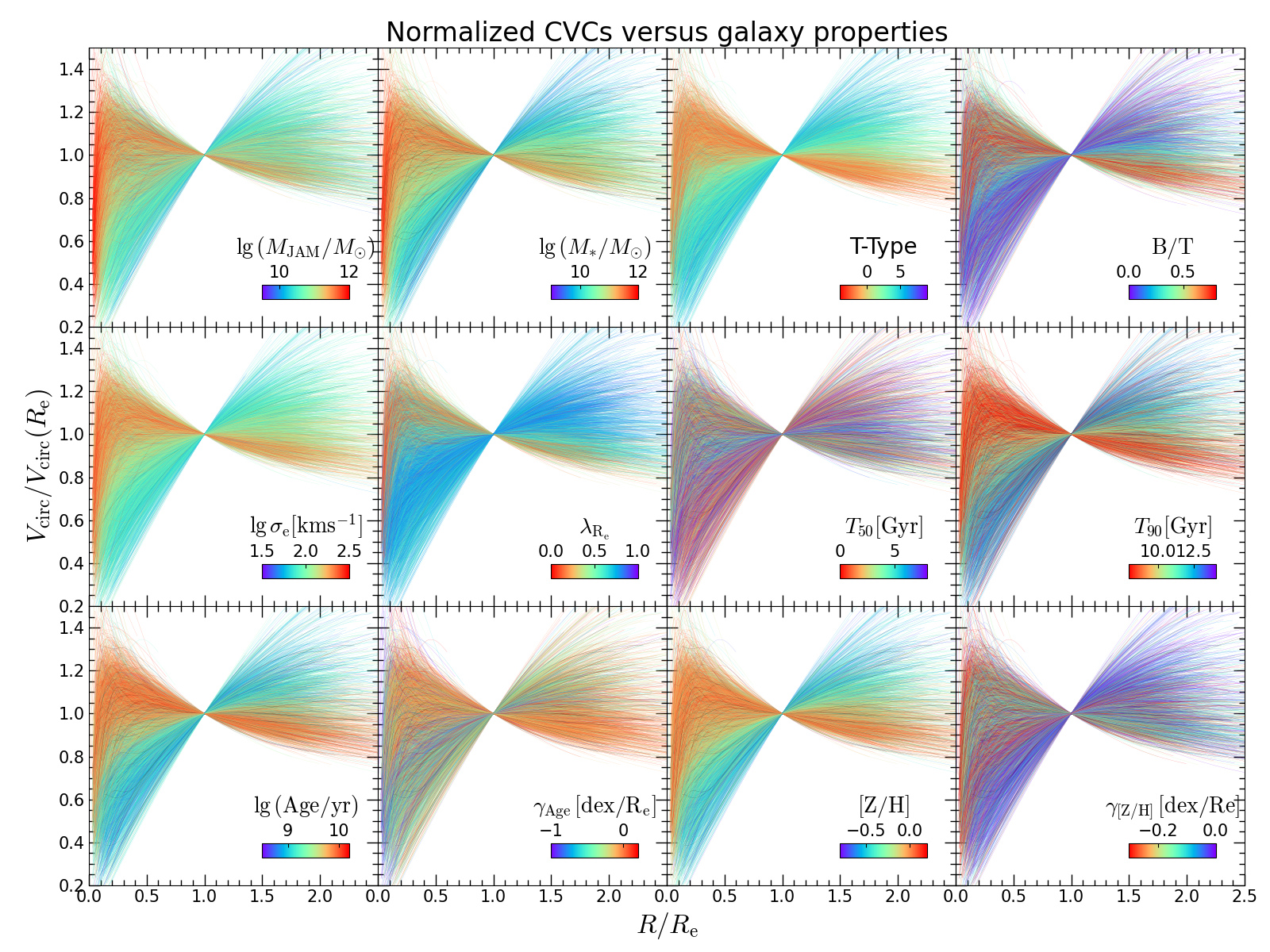}
    \caption{The normalized CVCs correlate with other galaxy properties. Definitions can be found in \autoref{subsec:Dyn} and \autoref{subsec:SPS}. These properties include masses (dynamical mass $M_{\rm JAM}$ and stellar mass $M_{\ast}$ based on Salpeter IMF), morphology (T-Type and $B/T$), velocity dispersion $\sigma_{\rm e}$, spin parameter $\lambda_{\rm R_{e}}$, the cosmic times when stars accounting for 50\% ($T_{50}$) and 90\% ($T_{90}$) of galaxies' present-day stellar mass were formed, luminosity-weighted stellar age, stellar metallicity, and their radial gradients.}
    \label{fig:Vcirc_galprop}
\end{figure*}

In \autoref{fig:Vcirc_galprop}, we present the normalized CVCs colored according to different galaxy properties. The top panels show a clear dependence of CVC shape on mass and morphology. Galaxies with higher mass (either dynamical mass $M_{\rm JAM}$ or stellar mass $M_{\ast}$), smaller T-Type values (indicating earlier types), and higher $B/T$ tend to have declining CVCs. In contrast, less massive, later-type galaxies with smaller $B/T$ tend to have rising CVCs. Flat CVCs at the outskirts are found in galaxies with intermediate mass and morphology (early-type spirals or S0), consistent with previous studies \citep{Kalinova2017,Yoon2021}.

The shape of CVCs also correlates with kinematic properties such as stellar velocity dispersion $\sigma_{\rm e}$ and the proxy for stellar angular momentum $\lambda_{\rm R_{e}}$ (first two panels in the second row of \autoref{fig:Vcirc_galprop}). As $\sigma_{\rm e}$ increases and $\lambda_{\rm R_{e}}$ decreases, trends transition from rising CVCs to flat, and finally to declining CVCs. These trends reflect differences in galaxy structure: $\sigma_{\rm e}$ traces the bulge fraction \citep{Cappellari2013b}, while $\lambda_{\rm R_{e}}$ values reflect the fraction of ordered motions, which is also related to the prominence of the disk component or bulge mass fraction.

The $T_{50}$ and $T_{90}$ values represent the formation times of stars (in cosmic time) that account for 50\% and 90\% of the galaxy's present-day stellar mass, respectively. The shape of CVCs shows clear dependence on $T_{90}$: declining CVCs for earlier-quenched galaxies (higher $T_{90}$) and rising CVCs for later-quenched galaxies (lower $T_{90}$). According to the two-phase evolutionary scenario on the mass-size plane \citep[as discussed in][]{Zhu2024}, galaxies increase their mass and size through gas accretion-induced \textit{in situ} star formation and dry mergers (by accreting \textit{ex situ} already formed stars). Earlier-quenched galaxies have already formed large bulges through \textit{in situ} star formation and continue morphological transformation through dry mergers, while later-quenched galaxies still undergo bulge growth and transformation into earlier types. In contrast, there is little or no dependence on $T_{50}$ because $T_{50}$ is less relevant for distinguishing between the two mechanisms.

We also examine the dependence of CVC shape on stellar age and metallicity, as well as their gradients, in the bottom panels of \autoref{fig:Vcirc_galprop}. Older and metal-rich galaxies with flatter age gradients tend to have declining CVCs. Conversely, rising CVCs usually appear in younger and metal-poor galaxies with negative age gradients. The weak dependence on metallicity gradients may relate to different trends of metallicity and metallicity gradients on the mass-size plane \citep[see figures 8 and 12 in][]{Lu2023}: metallicity follows velocity dispersion well, but the distribution of metallicity gradients is more complex.

In summary, our main findings are as follows:
\begin{enumerate}[label=(\roman*)]
\item Galaxies with declining CVCs tend to be massive, early-type, early-quenched, old, metal-rich, with high $B/T$, high velocity dispersion, low spin, and flat age gradients.
\item In contrast, galaxies with rising CVCs display opposite characteristics, while those with flat CVCs have features between declining and rising CVCs.
\item There is a weak dependence on $T_{50}$ and metallicity gradients.
\end{enumerate}

These quantities often relate to bulge and disk fractions, which are seen as the physical origins of the diverse CVC shapes. One might assume it is possible to predict CVC shape given bulge and disk fractions, and a DM fraction that accounts for the DM halo contribution. In the next subsection, we will assess whether this model (with a bulge, a disk, and a DM halo) can reproduce the observed diverse CVC shapes.

\subsection{A Bulge-Disk-Halo model to quantify the shape of CVCs}\label{subsec:BDH_model}

Based on the remarkable similarity of RCs first noted by \citet{Rubin1985}, there have been some studies trying to predict the amplitude and shape of RCs with a few key photometric parameters. For instance, \citet{Persic1996} and \citet{Salucci2007} proposed a universal rotation curve (URC) model for spiral galaxies, and \citet{Karukes2017} and \citet{DiPaolo2019} extended the URC model to dwarf and low-surface-brightness galaxies, respectively. Recently, \citet{Patel2024} developed a neural network (NN) model to predict the shape of RCs using the same photometric quantities as the literature URC, and found higher accuracy with their NN-based URC. They concluded that the improvement of literature URC at all radii requires detailed modeling in the inner region (including a bulge component) and at the outskirts (better parameterization of a DM halo). The inclusion of a bulge component is even more important when quantifying the shape of CVCs in MaNGA galaxies, given their substantial population of bulge-dominated systems that has been demonstrated through combined photometric and kinematic structural decompositions \citep{Rigamonti2023,Rigamonti2024}.

In this section, we try to use a bulge-disk-halo (hereafter BDH) model to predict the shape of CVCs. In this model, the bulge and disk components are described as two Sersic components, with the Sersic indices for the bulge and disk being $n_{\rm Ser, bulge}$ and $n_{\rm Ser, disk}=1$ (exponential), respectively, while the DM halo is assumed to be an NFW profile. For each Sersic component, the surface brightness is written as
\begin{equation}
    \Sigma(R)=\Sigma_{0}{\rm exp}\left[-b_{n}\left(\frac{R}{R_{e}}\right)^{1/n}\right],
\end{equation}
where $b_{n}$ is a function of Sersic index $n$, obtained by solving the equation \citep[rewritten from][equation~5]{Ciotti1991}
\begin{equation}\label{eq:beta_n}
    \frac{\Gamma(2n,b_{n})}{\Gamma(2n)}=Q(2n,b_{n})=\frac{1}{2},
\end{equation}
where $\Gamma(a)$ is the gamma function, $\Gamma(a,x)$ is the upper incomplete one, and $Q(a,x)$ is the regularized upper incomplete gamma function (\citealt{Olver2010}, \href{https://dlmf.nist.gov/8.2#E4}{equation~8.2.4}). Various approximations have been proposed to calculate $b_{n}$ \citep[e.g.,][]{Capaccioli1989,Prugniel1997,Ciotti1999,Cappellari2023}. However, one can also compute $b_n$ using the special function $Q^{-1}(a,s)$ giving the inverse of the regularized upper incomplete gamma function. This provides the solution for $z$ of $s=Q(a,z)$. From \autoref{eq:beta_n} one can then just write
\begin{equation}
    b_n = Q^{-1}(2n,1/2),
\end{equation}
The function $Q^{-1}(a,s)$ is implemented in popular languages like Python as \texttt{scipy.special.gammainccinv}, Mathematica as \texttt{InverseGammaRegularized}, or Matlab as \texttt{gammainccinv}. 

We adopt the analytical approximation in \citet{LimaNeto1999} to calculate the deprojected mass density of the Sersic model
\begin{equation}
    \rho(r) = \rho_{0}\left(\frac{r}{R_e}\right)^{-p_{n}}{\rm exp}\left[-b_{n}\left(\frac{r}{R_e}\right)^{1/n}\right],
\end{equation}
where $p_{n}$ is
\begin{equation}
    p_{n} = 1-\frac{0.594}{n}+\frac{0.055}{n^2},
\end{equation}
and the enclosed mass profile
\begin{equation}\label{eq:Sersic_M3d}
    M(r)=M_{\infty}\frac{\gamma[(3-p_{n})n, b_{n}(r/R_{e})^{1/n}]}{\Gamma[(3-p_{n})n]},
\end{equation}
where the enclosed mass at infinity (i.e. the total mass) is
\begin{equation}
    M_{\infty}=4\pi\rho_{0}R_{e}^3\,\frac{n\Gamma[(3-p_{n})n]}{b_{n}^{(3-p_{n})n}}.
\end{equation}
Readers are referred to \citet{Vitral2020} for a summary of different approximations for the deprojection of Sersic profile. For convenience, we assume a total stellar mass $M_{\rm \ast}$ and a projected half-light radius $R_{\rm e}$, which will finally be removed in the normalized CVC. We use the bulge-to-total ratio $B/T$, $R_{\rm e}$, and the $R_{\rm e, bulge}/R_{\rm e, disk}$ ratio to calculate $R_{\rm e, bulge}$ and $R_{\rm e, disk}$. Then we can obtain $M_{\rm \infty, bulge}=M_{\rm \ast}\times{\rm B/T}$ and $M_{\rm \infty, disk}=M_{\rm \ast}\times{\rm (1-B/T)}$ assuming a spatially constant stellar mass-to-light ratio, and further the $M_{\rm bulge}(r)$ and $M_{\rm disk}(r)$ profiles from \autoref{eq:Sersic_M3d}. Unlike previous studies that adopt different disk shapes depending on morphological types \citep[e.g. a spherical disk for ETGs and an infinitely thin disk for LTGs, as discussed in section~2.1 of][]{Dutton2011}, we assume a spherical disk for all galaxies. This assumption avoids introducing discontinuities by making the disk shape changing from spherical to infinitely thin dependent on morphologies. Although the spherical disk is a crude approximation, our toy model remains meaningful as it adopts various fitting functions of model parameters and does not pretend to be quantitatively accurate. Moreover, adopting an alternative extreme assumption that all galaxies have an infinitely thin disk \citep{Freeman1970} leads to a similar level of accuracy in our model (see \autoref{appendix:test_BDH} and \autoref{fig:cmp_JAM_BDH_Vcirc_InfThinDisk}), which justifies that the choice of a universal disk-shape assumption does not significantly affect the results. For the NFW DM halo with two free parameters ($r_{\rm s}$ and $\rho_{\rm s}$), we use the $r_{\rm s}/R_{\rm e}$ value to estimate $r_{\rm s}$ and use the DM fraction within a sphere of $R_{\rm e}$, defined as
\begin{equation}
    f_{\rm DM}(<R_{\rm e})=\frac{M_{\rm DM}(R_{\rm e})}{M_{\rm bulge}(R_{\rm e})+M_{\rm disk}(R_{\rm e})+M_{\rm DM}(R_{\rm e})},
\end{equation}
to calculate the $\rho_{\rm s}$ value of the NFW profile and finally obtain $M_{\rm DM}(r)$. The circular velocity curve 
\begin{equation}
    V_{\rm circ}(r)=\sqrt{\frac{G\times[M_{\rm bulge}(r)+M_{\rm disk}(r)+M_{\rm DM}(r)]}{r}}
\end{equation} 
is normalized by $R_{\rm e}$ and $V_{\rm circ}(R_{\rm e})$ to remove the dependence on the assumed $M_{\ast}$ and $R_{\rm e}$. To summarize, the normalized CVCs, $V_{\rm circ}(R)/V_{\rm circ}(R_{\rm e})$ versus $R/R_{\rm e}$\footnote{Throughout this paper, we use the circular velocity in the equatorial plane ($z=0$), which means $r=\sqrt{R^2+z^2}=R$.}, can be predicted using the five free parameters: $B/T$, $R_{\rm e, bulge}/R_{\rm e, disk}$, $n_{\rm Ser, bulge}$, $r_{\rm s}/R_{\rm e}$, and $f_{\rm DM}(<R_{\rm e})$. 

\begin{figure*}
    \centering
    \includegraphics[width=\textwidth]{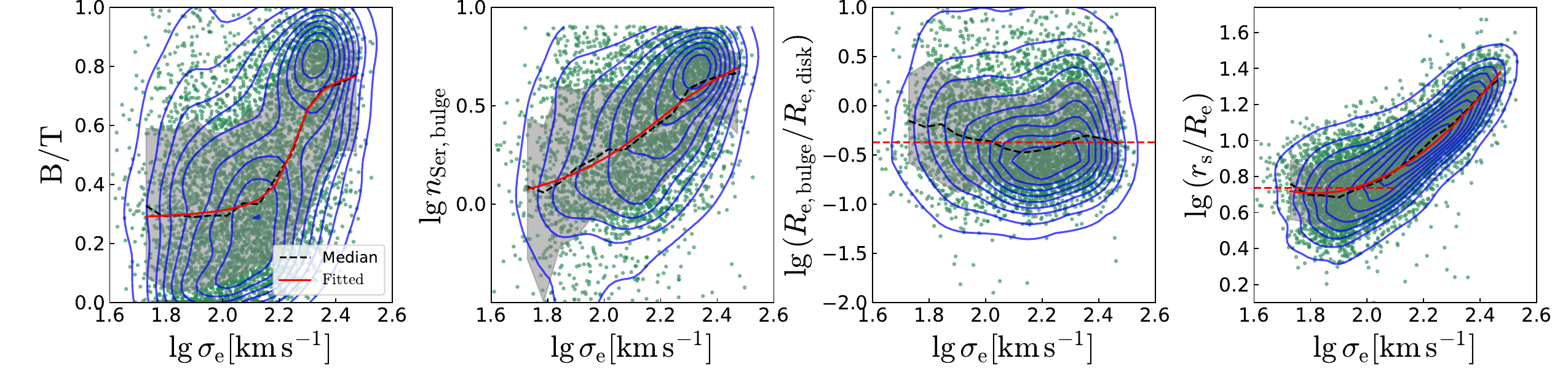}
    \caption{Scaling relations between structural parameters ($B/T$, $n_{\rm Ser, bulge}$, $R_{\rm e, bulge}/R_{\rm e, disk}$, $r_{\rm s}/R_{\rm e}$) and $\sigma_{\rm e}$ for the $\rm Qual\geqslant1$ galaxies. $B/T$, $n_{\rm Ser, bulge}$, and $R_{\rm e, bulge}/R_{\rm e, disk}$ are taken from \citet{Dominguez-Sanchez2022} (see \autoref{subsec:Dyn} for definitions). $r_{\rm s}/R_{\rm e}$ is the ratio between the dark halo scale radius and $R_{\rm e}$ of the whole galaxy, where $r_{\rm s}$ is estimated from the Chabrier IMF-based stellar mass, the stellar mass-to-halo mass relation in \citet{Moster2013}, and the mass-concentration relation in \citet{Dutton2014}. The blue contours are the kernel density estimate for the galaxy distribution, while the black dashed curve and the gray shaded region represent the median value and [16th, 84th] percentile of values, respectively. The best-fitting relations to the median trends (\autoref{eq:BT_sigma} and \autoref{eq:nser_sigma}) are shown in red solid curves. The horizontal red dashed line in the third panel represents the median value of $\lg\,(R_{\rm e, bulge}/R_{\rm e, disk})$ for the entire sample, while the horizontal red dashed line in the fourth panel represents the median value of $\lg\,(r_{\rm s}/R_{\rm e})$ for galaxies with $\lg\,(\sigma_{\rm e}/{\rm km\,s^{-1}})<2.1$.}
    \label{fig:BDH_scaling_relations}
\end{figure*}

Rather than predicting the CVC shape for individual galaxies, our aim is to use this simple toy model to predict the overall trends for specific galaxy populations. We use the scaling relations shown in \autoref{fig:BDH_scaling_relations}, along with the $f_{\rm DM}(<R_{\rm e})-\sigma_{\rm e}$ relation in \citet[][equation 7]{Lu2024}, to predict $B/T$, $n_{\rm Ser, bulge}$, $R_{\rm e, bulge}/R_{\rm e, disk}$, $r_{\rm s}/R_{\rm e}$, and $f_{\rm DM}(<R_{\rm e})$ for a given $\sigma_{\rm e}$. The best-fitting relations (red solid curves in \autoref{fig:BDH_scaling_relations}) for the median trends, assuming constant errors, are derived using arctan-based sigmoid functions:
\begin{gather}
    \label{eq:BT_sigma}
    B/T = 0.54-0.17\,\arctan\left[-16.37\,(\lg\sigma_{\rm e}-2.25)\right]\\
    \label{eq:nser_sigma}
    \lg n_{\rm Ser, bulge} = 0.43+0.37\,\arctan\left[3.08\,(\lg\sigma_{\rm e}-2.19)\right]
\end{gather}
The $B/T-\lg\,\sigma_{\rm e}$, $\lg\,n_{\rm Ser, bulge}-\lg\,\sigma_{\rm e}$, and $f_{\rm DM}(<R_{\rm e})-\lg\,\sigma_{\rm e}$ \citep[][figure~2]{Lu2024} relations show strong dependence on $\sigma_{\rm e}$, while $R_{\rm e, bulge}/R_{\rm e, disk}$ varies little with $\sigma_{\rm e}$. Although the median relation of $\lg\,(r_{\rm s}/R_{\rm e})-\lg\,\sigma_{\rm e}$ increases with higher $\sigma_{\rm e}$, however, the minimum $r_{\rm s}$ (for the median relation) is at least 5 times larger than $R_{\rm e}$, meaning the DM density profile within the kinematic data range is approximately power law. Thus we use the median values $\lg\,(R_{\rm e, bulge}/R_{\rm e, disk}) = -0.37$ and $\lg\,(r_{\rm s}/R_{\rm e})=0.74$ (the median value for $\lg\,(\sigma_{\rm e}/{\rm km\,s^{-1}})<2.1$, below which DM fractions become more significant) instead of the scaling relations for these two parameters.

\begin{figure*}
    \centering
    \includegraphics[width=\textwidth]{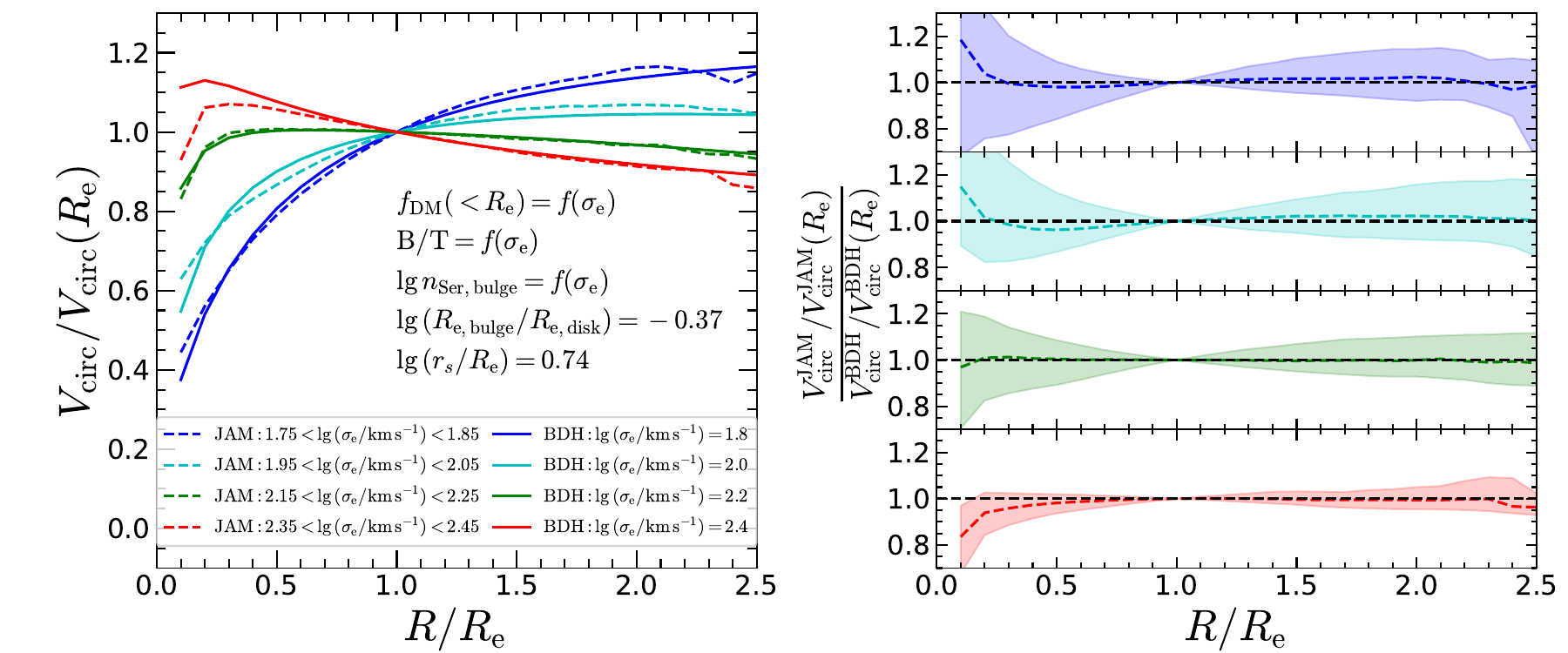}
    \caption{Comparisons between the BDH model-predicted shape of normalized CVCs (solid curves) and those derived from JAM (dashed curves) for four different velocity dispersion ($\sigma_{\rm e}$) bins, indicated by different colors. In the left panel, we show the median profile of the JAM-derived CVCs for each $\sigma_{\rm e}$ bin, while the BDH model predictions are calculated at the bin centers using the scaling relations (three relations as a function of $\sigma_{\rm e}$, and two are constants) presented in the left panel. The ratio between the JAM-derived and BDH-predicted CVCs, along with the associated error (represented by the [16th, 84th] percentiles of the JAM-derived CVCs), is shown in the right panel.}
    \label{fig:cmp_JAM_BDH_Vcirc}
\end{figure*}

\autoref{fig:cmp_JAM_BDH_Vcirc} shows the comparisons between CVC shapes predicted by the BDH model and those derived from JAM. As $\sigma_{\rm e}$ increases, galaxies' $f_{\rm DM}(<R_{\rm e})$ decreases, $B/T$ increases, and $n_{\rm Ser, bulge}$ becomes higher, the shape of CVCs changes from rising (blue curve) to flat (cyan curve) and then declining (red curve). A remarkably small systematic bias can be seen for different $\sigma_{\rm e}$ bins (with a bin width of 0.1 dex), in which we compare the median profile (dashed) of JAM-derived CVCs and the BDH-predicted CVC (solid). This indicates that the physical origin of the diverse CVC shapes is the relative contribution (both the mass ratios and scale radii) of different components (reflected by the scaling relations above), and the dependence on stellar mass found in previous studies is merely due to those relations being dependent on stellar mass. Among these relations, the two most important are $B/T$ and $f_{\rm DM}(<R_{\rm e})$, while the $n_{\rm Ser, bulge}$ is less important (comparing \autoref{fig:cmp_JAM_BDH_Vcirc} and \autoref{fig:cmp_JAM_BDH_Vcirc_fixednser}). 

An interesting fact is that, given a velocity dispersion $\sigma_{\rm e}$, we can estimate the amplitude $V_{\rm circ}(R_{\rm e}^{\rm maj})$ (or $V_{\rm circ}^{\rm max}$) of CVC with an error of 9\% (or 8\%) and predict the CVC shape with nearly zero bias in most cases. This demonstrates that $\sigma_{\rm e}$ is a good proxy for the inner gravitational potential. In particular, when IFU observations for high-redshift galaxies are quite expensive, one can apply the empirical aperture correction for velocity dispersion \citep{Zhu2023b} to obtain $\sigma_{\rm e}$ as the proxy for inner gravitational potential.

\begin{figure*}
    \centering
    \includegraphics[width=\textwidth]{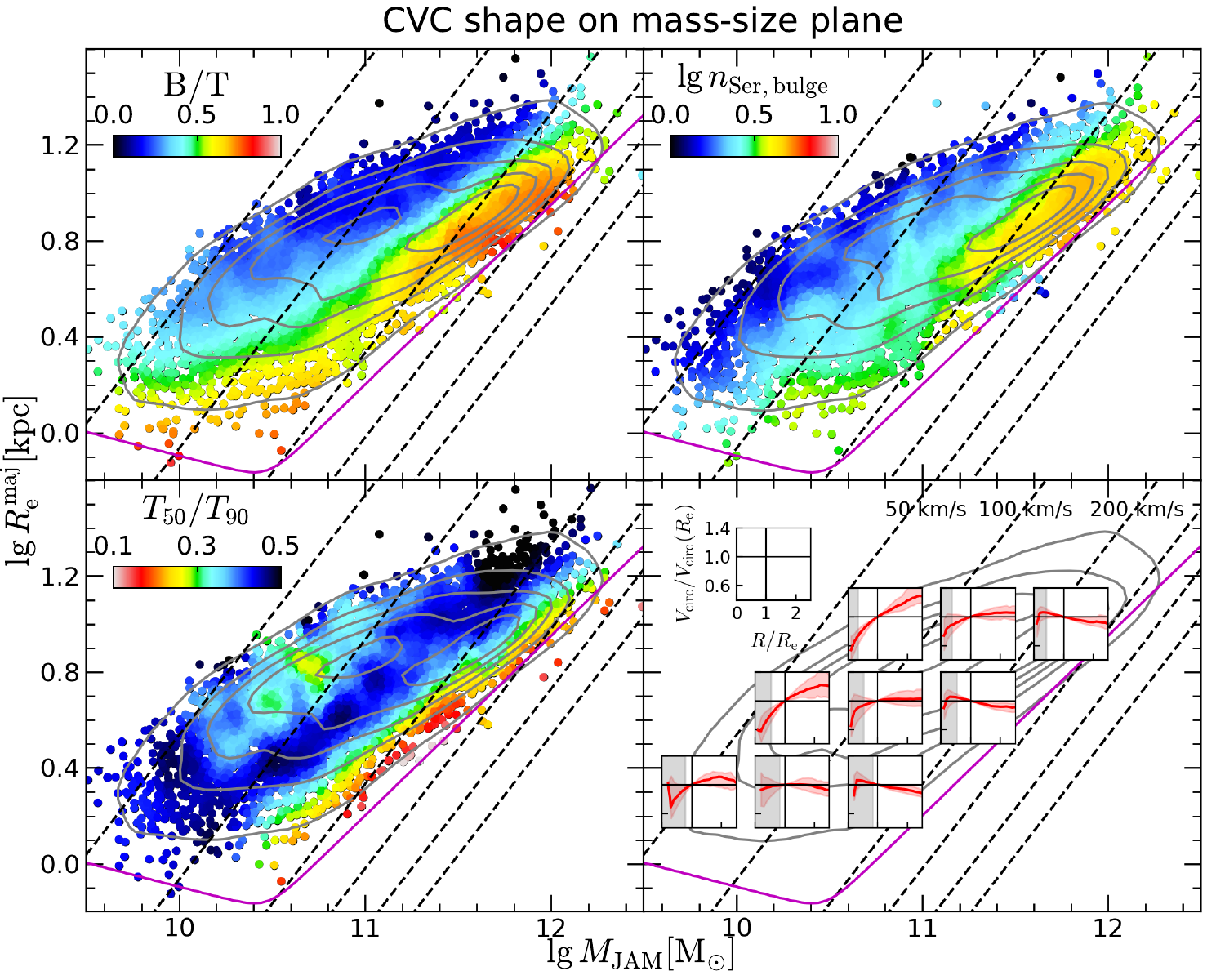}
    \caption{The distributions of $B/T$ (top-left panel), $\lg\,n_{\rm Ser, bulge}$ (top-right panel), $T_{50}/T_{90}$ (bottom-left panel), and the median CVC within a given mass and size bin (bottom-right panel) are shown on the $M_{\rm JAM}-R_{\rm e}^{\rm maj}$ plane. Distributions of $f_{\rm DM}(<R_{\rm e})$ can be found in \citet[][top-right panel of figure~16]{Zhu2024}. The distributions are smoothed by the \textsc{loess} software \citep{Cappellari2013b} with \texttt{frac=0.05}. In all panels, the dashed lines correspond to effective velocity dispersion $\sigma_{\rm e}=$ 50, 100, 200, 300, 400, and 500 $\rm km\,s^{-1}$ from left to right, which are calculated using the scalar virial equation $\sigma_\mathrm{e}^2 \equiv G M_{\rm JAM}/(5 R_{\rm e}^{\rm maj})$ (with factor 5 from \citealt{Cappellari2006}). The magenta curve shows the zone of exclusion (ZOE) defined in \citep[equation~4]{Cappellari2013b}, with the ZOE above $M_{\rm JAM}=2\times10^{10}{\rm M_{\odot}}$ is approximately $R_{\rm e}^{\rm maj}\propto M_{\rm JAM}^{0.75}$. The gray contours show the kernel density estimate for the galaxy distribution. \textit{Bottom-right panel}: The inset panels in the bottom-right panel show the median (red solid) and the [16th, 84th] percentiles (red shaded region), as well as the median FWHM of PSF (vertical shaded gray region). All the inset panels have the same physical scales as the bottom-right panel, while the ticks and labels are shown in the upper-left inset panel.}
    \label{fig:mass_size_Vcirc_shape}
\end{figure*}

\subsection{CVC shape on the mass-size plane}\label{subsec:CVCshape_mass_size}

It has been suggested by the IFU results of nearby ETGs \citep{Cappellari2013b,Cappellari2016,Zhu2024} and the observations of high-redshift ETGs \citep{vanderWel2008,vanDokkum2015,Derkenne2021} that galaxy evolution on the mass-size plane follows a simple two-phase scenario \citep{Oser2010,Tonini2016}: (i) \textit{in situ} star formation and (ii) \textit{ex situ} accretion by dry mergers (minor or major). \textit{In situ} star formation, whether triggered by infalling cold gas (e.g. cold streams or gas-rich minor mergers) or pure secular evolution without gas replenishment, mildly increases galaxy sizes when building up stellar masses \citep{Oser2010}. In contrast, dry major mergers lead to proportional increases in both galaxy sizes and masses (while maintaining nearly constant $\sigma_{e}$), whereas dry minor mergers result in more significant size growth, typically by a factor of 4 for mass doubling \citep{Naab2009}.

The two-phase scenario is not intended as a full picture of galaxy formation but instead serves as a framework for understanding the physical processes in a straightforward way. As discussed in previous sections, the shape of CVCs is determined by $B/T$, $f_{\rm DM}(<R_{\rm e})$, and $n_{\rm Ser, bulge}$. Consequently, the evolution of CVC shapes on the mass-size plane is related to the evolution of these parameters. Since the distribution of $f_{\rm DM}(<R_{\rm e})$ has been presented in \citet[][figure~16]{Zhu2024}, we show the distributions of $B/T$, $n_{\rm Ser, bulge}$, $T_{50}/T_{90}$, and the median CVC for galaxies in each mass-size bin (corresponding to each inset panel) in \autoref{fig:mass_size_Vcirc_shape}. $T_{50}/T_{90}$ is the ratio between cosmic times when 50\% and 90\% stellar masses are formed. The bottom-right panel reveals clear trends of CVC shapes:
\begin{itemize}
    \item Moving from left to right on the mass-size plane, the CVC shapes transition from rising in lower-mass galaxies to declining in higher-mass galaxies.
    \item At fixed mass, galaxies with larger size tend to have rising CVC shapes, while smaller galaxies typically exhibit declining CVCs.
    \item Galaxies closest and parallel to the zone of exclusion (ZOE; $R_{\rm e}^{\rm maj}\propto M_{\rm JAM}^{0.75}$), which have the highest velocity dispersion of $\sigma_{\rm e}\approx200\,{\rm km\,s^{-1}}$ (see the rightmost three inset panels), maintain declining CVCs, with little change in shape as mass and size increase.
\end{itemize}

When disk galaxies form \citep[i.e., the formation of stellar disks;][]{Mo1998}, they typically exhibit small bulges (low $\sigma_{\rm e}$ and low $B/T$). These young disk galaxies tend to host pseudobulges \citep[e.g.][]{Hu2024}, which are induced by secular evolution and typically exhibit exponential-like profiles with Sersic indices of $n_{\rm Ser, bulge} \approx 1-2$ \citep{Kormendy2004}. Inner \textit{in situ} star formation, whether driven by secular evolution or enhanced through gas accretion, drives bulge growth, resulting in a slightly higher Sersic index for the bulge and a lower central DM fraction. This transformation causes galaxies to shift their CVC shapes from rising to declining as their bulges and $\sigma_{\rm e}$ increase. The size dependence (quantified by the half-light radius) is primarily explained by the fact that, at fixed mass, galaxies with higher $B/T$ are smaller, and contain less DM within their smaller size (lower DM fraction). The trend differs for high-$\sigma_{\rm e}$ ETGs that lie parallel to the ZOE, where dry mergers dominate mass and size growth. Along the direction of the ZOE, dry mergers mildly increase $B/T$ and have minimal impact on $f_{\rm DM}(<R_{\rm e})$. Although $n_{\rm Ser, bulge}$ changes more significantly, it plays a relatively less important role in determining CVC shape (see \autoref{subsec:BDH_model}). 
As a result, these high-$\sigma_{\rm e}$ galaxies exhibit nearly the same CVC shape.

In the bottom-left panel of \autoref{fig:mass_size_Vcirc_shape}, the star-forming galaxies at $10^{10}-10^{11}{\rm M_{\odot}}$ exhibit relatively smaller $T_{50}/T_{90}$ values compared to other star-forming galaxies with lower or higher masses. This is likely an \textit{in situ} star formation region dominated by pure secular evolution. Among the \textit{in situ} star formation processes, secular evolution dominates at intermediate masses ($10^{10}-10^{11}{\rm M_{\odot}}$), as gas accretion rates/merger rates are typically higher at lower/higher masses \citep{Hopkins2010,Tonini2016}. In the absence of gas replenishment, secular evolution tends to form stars at progressively slower rates (due to declining gas densities), resulting in the observed smaller $T_{50}/T_{90}$ ratios.

Other mechanisms are also potentially responsible for the evolution of galaxy mass and size, such as tidal heating and tidal stripping. Tidal heating puffs up the galaxies to form ultra-diffuse galaxies \citep{Jones2021,Fielder2024}, while tidal stripping leads to ultra compact dwarfs \citep{Penny2014,Mayes2021}. However, these environmental effects only work for low-mass satellite galaxies in a dense environment, which is not suitable for our sample dominated by central galaxies \citep[][section~2.5]{Zhu2024}.

\section{Conclusions}\label{sec:conlusions}
In the seventh paper of the MaNGA DynPop project, we derive the CVCs (or intrinsic RCs) for 6000 nearby galaxies from stellar dynamical models in \citep{Zhu2023}. The amplitude and shape of CVCs (corrected for the inclination angle and seeing effects) are closely related to the inner gravitational potential of galaxies, providing key insights into galaxy formation and evolution scenarios. By combining these with spatially resolved stellar population properties \citep{Lu2023}, we study the scaling relations between CVCs (shape and amplitude) and other galaxy properties. We propose a simple BDH model that includes a bulge, a disk, and a DM halo to predict the shape of CVCs across different galaxy types.

The main conclusions are summarized below.

\begin{itemize}
    \item The amplitude of CVCs, characterized by the circular velocity at the half-light radius, $V_{\rm circ}(R_{\rm e}^{\rm maj})$, or the maximum circular velocity within the FoV, $V_{\rm circ}^{\rm max}$, is linearly related to the velocity dispersion within one effective radius $\sigma_{\rm e}$, linking the Tully-Fisher relation \citep{Tully1977} and the Faber-Jackson relation \citep{Faber1976}. The best-fitting relations are $V_{\rm circ}(R_{\rm e}^{\rm maj})\approx1.62\times\sigma_{\rm e}$ and $V_{\rm circ}^{\rm max}\approx1.72\times\sigma_{\rm e}$ with a small error of 7\% (\autoref{fig:VcRmaj_Vmax_sigma}), respectively.
    \item The shape of CVCs (rising, flat, and declining) correlates with galaxy dynamical and stellar population properties: galaxies with declining CVCs are massive, early-type, early-quenched, old, metal-rich, and exhibit high $B/T$, low spin, high velocity dispersion, and a flat age gradient, while galaxies with rising CVCs exhibit the opposite properties (\autoref{fig:Vcirc_galprop}).
    \item We propose a BDH model, which includes a Sersic bulge, an exponential disk, and a NFW DM halo, to quantify the shape of CVCs. In this model, given the  bulge-to-total ratio $B/T$, the dark matter fraction within $f_{\rm DM}(<R_{\rm e})$, the Sersic index of the bulge component $n_{\rm Ser, bulge}$, the ratio between the bulge and disk effective radii $R_{\rm e, bulge}/R_{\rm e, disk}$, and the ratio between the scale radius of NFW halo and the luminous half-light radius $r_{\rm s}/R_{\rm e}$, one can predict the shape of CVCs. We test this model for galaxies within different $\sigma_{\rm e}$ bins, in which the free parameters are predicted from their empirical scaling relations with $\sigma_{\rm e}$ \citep[see \autoref{fig:BDH_scaling_relations} and figure~2 in][]{Lu2024}, finding a nearly unbiased consistency with JAM-derived CVC shapes (\autoref{fig:cmp_JAM_BDH_Vcirc}). The BDH model quantitatively confirms that the shape of CVCs is mainly driven by $B/T$, $f_{\rm DM}(R_{\rm e)}$, and $n_{\rm Ser, bulge}$: galaxies with a larger $B/T$, lower $f_{\rm DM}(R_{\rm e})$, and larger $n_{\rm Ser, bulge}$ tend to have declining CVCs, while rising CVCs are found in galaxies with the opposite properties.
    \item The evolution of CVC shapes on the mass-size plane is closely related to the evolution of $B/T$, $f_{\rm DM}(R_{\rm e})$, and $n_{\rm Ser, bulge}$, supporting the scenario of two evolutionary channels \citep[e.g.][]{Cappellari2013b,vanDokkum2015,Zhu2024}: (i) \textit{in situ} star formation (through gas accretion, gas-rich minor mergers, or secular evolution) moving galaxies from left to right (CVC shape from rising to declining), inducing bulge growth, reducing the central DM fraction, and increasing the Sersic index of bulges; (ii) dry mergers moving gas-poor galaxies along the constant $\sigma_{\rm e}$ lines upwards, mildly increasing $B/T$ and the Sersic index of bulges, leaving the central DM fraction unchanged, and finally leading to the same CVC shapes along the direction of the ZOE (\autoref{fig:mass_size_Vcirc_shape}).
\end{itemize}

We will release all data derived from the stellar dynamical models, including 3D mass distributions (mass density profiles), 2D mass distributions (surface mass density maps), and circular velocity curves. One can use the MGE coefficients and best-fitting free parameters taken from the catalog \citep{Zhu2023} to derive the mass distributions of stellar, DM, and total components. The database and a Python script for the calculation will be provided on the website of MaNGA DynPop (\url{https:manga-dynpop.github.io}).

\begin{acknowledgments}

We thank the anonymous referee for the very useful comments. This work is supported by the National Science Foundation of China (Grant No. 11821303 to S.M.). K.Z. acknowledges the support from the Shuimu Tsinghua Scholar Program of Tsinghua University.

Funding for the Sloan Digital Sky 
Survey IV has been provided by the 
Alfred P. Sloan Foundation, the U.S. 
Department of Energy Office of 
Science, and the Participating 
Institutions. 

SDSS-IV acknowledges support and 
resources from the Center for High 
Performance Computing  at the 
University of Utah. The SDSS 
website is www.sdss.org.

SDSS-IV is managed by the 
Astrophysical Research Consortium 
for the Participating Institutions 
of the SDSS Collaboration including 
the Brazilian Participation Group, 
the Carnegie Institution for Science, 
Carnegie Mellon University, Center for 
Astrophysics | Harvard \& 
Smithsonian, the Chilean Participation 
Group, the French Participation Group, 
Instituto de Astrof\'isica de 
Canarias, The Johns Hopkins 
University, Kavli Institute for the 
Physics and Mathematics of the 
Universe (IPMU) / University of 
Tokyo, the Korean Participation Group, 
Lawrence Berkeley National Laboratory, 
Leibniz Institut f\"ur Astrophysik 
Potsdam (AIP),  Max-Planck-Institut 
f\"ur Astronomie (MPIA Heidelberg), 
Max-Planck-Institut f\"ur 
Astrophysik (MPA Garching), 
Max-Planck-Institut f\"ur 
Extraterrestrische Physik (MPE), 
National Astronomical Observatories of 
China, New Mexico State University, 
New York University, University of 
Notre Dame, Observat\'ario 
Nacional / MCTI, The Ohio State 
University, Pennsylvania State 
University, Shanghai 
Astronomical Observatory, United 
Kingdom Participation Group, 
Universidad Nacional Aut\'onoma 
de M\'exico, University of Arizona, 
University of Colorado Boulder, 
University of Oxford, University of 
Portsmouth, University of Utah, 
University of Virginia, University 
of Washington, University of 
Wisconsin, Vanderbilt University, 
and Yale University.
\end{acknowledgments}

\appendix

\section{Dependence of the BDH model on the assumptions}\label{appendix:test_BDH}
\begin{figure*}
    \centering
    \includegraphics[width=\textwidth]{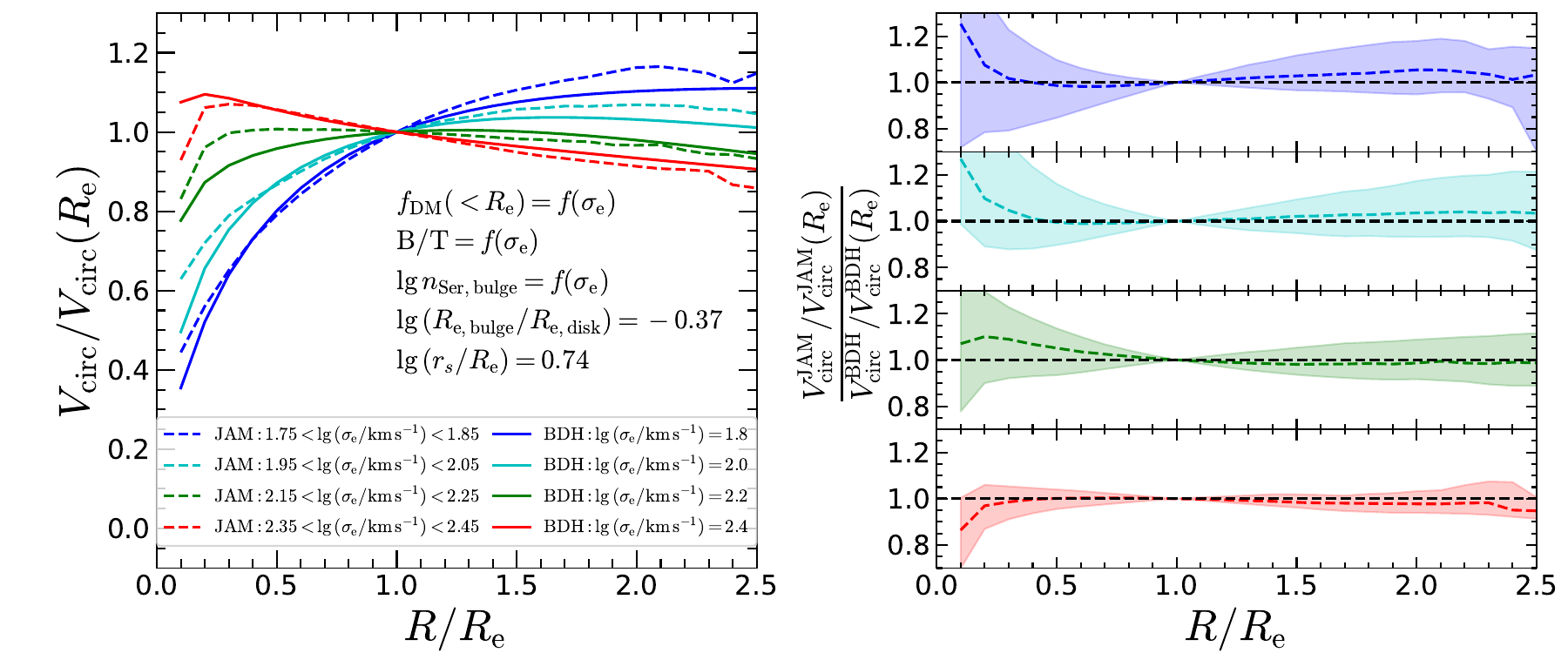}
    \caption{Similar to \autoref{fig:cmp_JAM_BDH_Vcirc}, but adopting an alternative extreme assumption that all galaxies have infinitely thin exponential disks \citep{Freeman1970}.}
    \label{fig:cmp_JAM_BDH_Vcirc_InfThinDisk}
\end{figure*}

\begin{figure*}
    \centering
    \includegraphics[width=\textwidth]{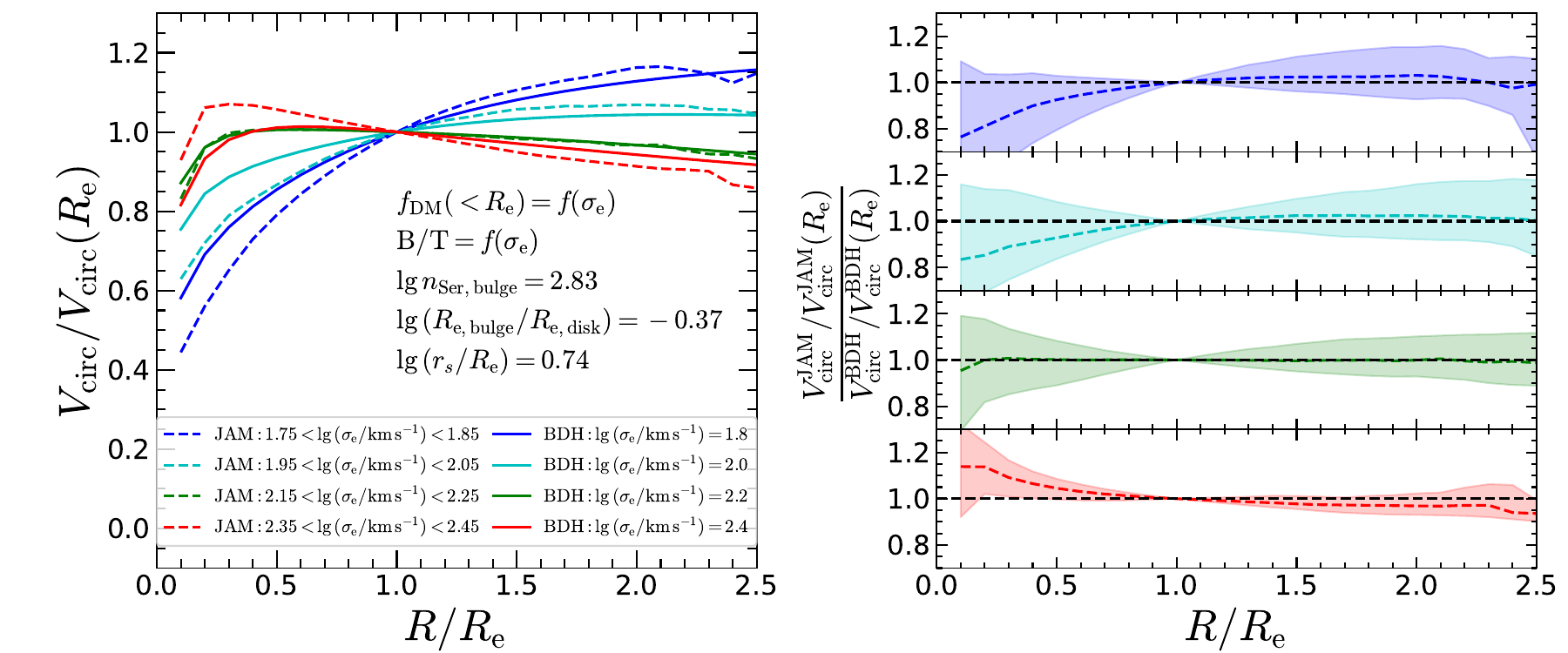}
    \caption{Similar to \autoref{fig:cmp_JAM_BDH_Vcirc}, but using a fixed $n_{\rm Ser, bulge}=2.83$ that corresponds to a Hernquist profile in the inner region \citep[Appendix~A and figure~A1 in][]{Vitral2020}. The trends of the CVC shape changing with $\sigma_{\rm e}$ still exist, but the BDH model-predicted curves tend to overestimate at the low $\sigma_{\rm e}$ end and underestimate at the high $\sigma_{\rm e}$ end.}
    \label{fig:cmp_JAM_BDH_Vcirc_fixednser}
\end{figure*}

We adopt a spherical exponential disk assumption, $V_{\rm disk}(r)=\sqrt{GM_{\rm disk}(r)/r}$, in the standard BDH model (\autoref{fig:cmp_JAM_BDH_Vcirc}), which might not be a good approximation for LTGs. Here we test if the adopted assumptions affect the results. In \autoref{fig:cmp_JAM_BDH_Vcirc_InfThinDisk}, we adopt the infinitely thin disk \citep{Freeman1970} assumption for all galaxies (including ETGs and LTGs). Under this alternative extreme assumption, the circular velocity at radius $r$ (in the plane of the disk) contributed from the disk component is 
\begin{equation}
V_{\rm disk}(r)=\frac{GM_{\rm \infty, disk}}{R_{\rm d}}2y^{2}\left[I_{0}(y)K_{0}(y)-I_{1}(y)K_{1}(y)\right],
\end{equation}
where $M_{\rm \infty, disk}$ is the total mass of the disk, $R_{\rm d}=R_{\rm e, disk}/1.678$ is the disk scale length, $y=(r/R_{\rm d})$, and $I_{\rm n}$ and $K_{\rm n}$ are the modified Bessel functions \citep{Freeman1970}. We find the predicted normalized $V_{\rm circ}$ profiles under the two extreme assumptions (spherical exponential disk or thin exponential disk) are nearly identical. In \autoref{fig:cmp_JAM_BDH_Vcirc_fixednser}, we fix the Sersic index of the bulge component to be 2.83, which is similar to the \citet{Hernquist1990} profile \citep[see Appendix~A and figure~A1 in][]{Vitral2020}, and find that adopting a universal Hernquist bulge can roughly predict the trend of CVCs (rising, flat or declining) but can not quantitatively reproduce the shape of CVCs in the inner region.

\bibliography{ref}{}
\bibliographystyle{aasjournal}

\end{document}